OBSERVER-SIDE PARAMETER ESTIMATION FOR ADAPTIVE CONTROL

By

JASON NEZVADOVITZ

A THESIS PRESENTED TO THE GRADUATE SCHOOL
OF THE UNIVERSITY OF FLORIDA IN PARTIAL FULFILLMENT
OF THE REQUIREMENTS FOR THE DEGREE OF
MASTER OF SCIENCE

UNIVERSITY OF FLORIDA

2017

ACKNOWLEDGMENTS

I'd like to thank my research advisor, Dr. Dixon, for introducing me to the interesting field of adaptive control. I'd also like to thank my good friend, Forrest Voight, for listening to all my crazy ideas and providing insightful feedback. And finally, great thanks to my parents for supporting me in my pursuit of higher education.

TABLE OF CONTENTS

LIST OF FIGURES

Abstract of Thesis Presented to the Graduate School
of the University of Florida in Partial Fulfillment of the
Requirements for the Degree of Master of Science

OBSERVER-SIDE PARAMETER ESTIMATION FOR ADAPTIVE CONTROL

By

Jason Nezvadovitz

August 2017

Chair: Warren Dixon
Major: Mechanical Engineering

In adaptive control, a controller is precisely designed for a certain model of the system, but that model's parameters are updated online by another mechanism called the adaptive update. This allows the controller to aim for the benefits of exact model knowledge while simultaneously remaining robust to model uncertainty.

Like most nonlinear controllers, adaptive controllers are often designed and analyzed under the assumption of deterministic full state feedback. However, doing so inherently decouples the adaptive update mechanism from the probabilistic information provided by modern state observers.

The simplest way to reconcile this is to let the observer produce both state estimates and model parameter estimates, so that all probabilistic information is shared within the framework of the observer. While this technique is becoming increasingly common, it is still not widely accepted due to a lack of general closed-loop stability proofs.

In this thesis, we will investigate observer-side parameter estimation for adaptive control by precisely juxtaposing its mechanics against the current, widely accepted adaptive control designs. Additionally, we will propose a new technique that increases the robustness of observer-based adaptive control by following the same line of reasoning used for the popular "concurrent learning" method.

# CHAPTER 1
# INTRODUCTION

Typical robust control design first assumes known bounds for all system uncertainties (e.g. parameters, disturbances), and then prescribes a feedback loop that will stabilize the system's tracking error as long as said uncertainty bounds are respected.

The looser the bounds, the stronger the feedback must be to dominate the effects of the large uncertainties, thus increasing nominal actuator effort and amplification of noise. If tighter bounds are known, the controller can be more finely tuned to improve transient response and steady-state error. However, acting on tight uncertainty bounds increases the risk of unexpected performance and even instability [1].

This trade-off is what motivates adaptive control. In adaptive control, a controller is precisely constructed for a certain model of the system, but that model is updated online by another feedback loop called the adaptive update. This allows the controller to aim for the benefits of exact model knowledge while simultaneously remaining robust to model uncertainty.

The adaptive update can be thought of as a system identification, but in actuality an adaptive controller need not accurately identify the system. Many simple adaptive updates depend only on the tracking error, so as the tracking error approaches zero, the model estimate settles. Therefore, the model estimate will only be accurate if the current trajectory being tracked excites all modes of the model, a somewhat impractical condition known as persistence of excitation [2].

For some designers, obtaining an accurate model is extraneous. However, it does have performance benefits: when a less common mode of the dynamic is excited, but an accurate model is already known, extraordinary transient performance can be achieved. Additionally, an accurate model can be used for important related computations like path planning. Consequently, adaptive control with practical, accurate system identification has been a major focus of research in the last few decades [3].

The obvious way to achieve accurate system identification without persistence of excitation is to have the adaptive update depend on the model error itself, i.e. the difference between the true model values and the estimated model values. Of course, the model error cannot be computed directly (the true model is not known!), so most methods work with a related quantity called the prediction error [4].

The prediction error is the difference between what the system did and what the model estimate said it would do. If the system's full state is known exactly, then all of the prediction error can be attributed to model error. This relationship warrants its substitution into the adaptive update.

However, perfect measurements of the state are never really available. Instead, a set of noisy sensors provides measurements related to the state, and the state is estimated by a feedback loop on sensor prediction error. This loop is called the observer.

Whether it's a transient or a long-term bias, any inaccuracy in the observer's state estimate becomes part of the prediction error, thus skewing its use as a pure indicator of model error in the adaptive update.

It isn't surprising that this issue is hardly addressed in the literature. Nonlinear control results often assume perfect state estimation (e.g. [3] and [5]) because the lack of a separation principle and the need for stochastic analysis make stability proofs immensely more complicated. Some results like [6] do consider an observer in the loop, but they address the special case of deterministic output-feedback, in which the quantity to be controlled has perfect measurements. In many applications (especially robotics), the quantity to be controlled *is* the full state, so output-feedback still implies perfect state estimation. To compensate, strong results come paired with experiments that demonstrate their algorithm's success in the real world, where noisy output sensors and a stochastic observer are used, even though the mathematical development does not consider them (e.g. [7]).

Putting aside the theoretical issues with that justification, there are still practical questions raised. Stochastic observers (like the popular Kalman filter) provide an estimate of both the state and its probability distribution [8], but the latter is effectively thrown away by the controller. Couldn't the adaptive update use that information to help distinguish model error from state estimation error? Even if an ordinary adaptive control design is proven to be robust to bounded state estimation error, certainly transient performance can be improved by utilizing the probabilistic state information.

For example, if the state estimate probability distribution expresses growing or large uncertainty in a particular state, then perhaps the adaptive update should slow-down learning for the parameters closely coupled to that state to avoid its erroneous influence. In fact, this is essentially what a stochastic observer like the Kalman filter already does: the state estimate probability distribution weights the causes of sensor prediction error to help the corrective action reject noise.

In this light, it's somewhat bizarre that the adaptive update is not just a part of the observer; their tasks are very similar. If the model uncertainties are treated as additional states, and it is shown that they are observable, then an observer can estimate them just fine. Furthermore, by bringing model estimation over to the observer-side, no new mechanism is needed to have the model estimation make use of the probabilistic state information; it will all be shared under the general framework of the stochastic observer.

Indeed, this is consistently done in practice, typically using either an extended Kalman filter (EKF), unscented Kalman filter (UKF), or particle filter to simultaneously estimate both the state and the model (see [9], [10], and [11] respectively). These stochastic observers have many convergence proofs (e.g. [12] and [13]), but unfortunately, stability proofs are rare for the full closed-loop (i.e. when the model estimates feed into a control law). This is partly due to the same reason that common adaptive control results don't consider stochastic state estimation: the combined analysis is very complicated for arbitrary nonlinear systems.

However, the real reason might be that there just isn't a general stability proof. If the observer doesn't care at all about tracking error, then how can one guarantee that the parameter estimates do not pass through a configuration that leads to controller instability during the observer's convergence? Well, the same question can be asked about the state estimates too! Theorists seem to be less concerned about the latter in adaptive control publications just because effective state estimation has become so common.

The purpose of this thesis is to investigate observer-side parameter estimation for adaptive control, by precisely juxtaposing its mechanics against the current, widely accepted adaptive control designs. Additionally, we will propose a new technique that increases the robustness of observer-based adaptive control by following the same line of reasoning used for concurrent learning in [3].

# CHAPTER 2
# THE ESSENTIAL ADAPTIVE CONTROL LAW

Nonlinear, time-invariant, control-affine dynamics model a very wide range of physical systems and are usually the focus of adaptive control results. For a system state vector $x(t) \in \mathbb{R}^n$ and a control-input vector $u \in \mathbb{R}^m$, the dynamics are,

$$\dot{x} = f(x) + B(x)u$$

where $f : \mathbb{R}^n \to \mathbb{R}^n$ is the drift dynamic and $B : \mathbb{R}^n \to \mathbb{R}^{n \times m}$ is the actuator dynamic. The system is assumed to be at least fully-actuated, meaning that $B(x)$ is always full row rank (or more generally, that the actuator dynamic term spans the entire state space).

We desire $x(t)$ to track some reference state $r(t) \in \mathbb{R}^n$, so we define a tracking error $e$ and examine its dynamics,

$$e := r - x$$

$$\dot{e} = \dot{r} - f(x) - B(x)u$$

We typically desire exponential convergence of the error to zero, not just because exponential decay is fast, but also because its relationship with linear systems will allow us to evaluate the closed-loop performance with well-understood linear stability metrics like gain and phase margin [3]. Thus, for some $K > 0$, we want to choose $u$ so that,

$$\dot{e} \stackrel{\text{want}}{=} -Ke$$

Achieving this linear closed-loop error dynamic is called feedback linearization, and it is the core of most adaptive control laws. Because $r$ is user-selected, $\dot{r}$ is known exactly. If $x$ and the functions $f$ and $B$ are also known exactly, then we can easily choose $u$ as,

$$u \stackrel{\text{ideal}}{:=} B^+(x)\Big(Ke + \dot{r} - f(x)\Big)$$

where $B^+ = B^T(BB^T)^{-1}$ is a pseudoinverse of $B(x)$ that satisfies $B(x)B^+(x) = I \ \forall x$ [14]. (The fully-actuated assumption guarantees that $BB^T$ is invertible).

Unfortunately, we only have *estimates* of the state, drift dynamic, and actuator dynamic, denoted as $\hat{x}$, $\hat{f}$, and $\hat{B}$ respectively. (We will also denote $\hat{e} := r - \hat{x}$). In hopes that our estimates are "correct enough," we use them anyway, choosing $u$ as

$$u := \hat{B}^+(\hat{x})\Big(K\hat{e} + \dot{r} - \hat{f}(\hat{x})\Big)$$

This control law is deemed "adaptive" when $\hat{f}$ and/or $\hat{B}$ are varied online to guarantee stability and improve performance. (The need to vary $\hat{x}$ online to track $x$ is a far more fundamental and obvious requirement, usually distinguished as "the observer's job").

In many applications, $B$ is actually known with good precision. After all, the actuators are usually already-engineered components, sometimes with their own internal control systems. Moreover, uncertainties that multiply the control-input ("unmatched" uncertainties) are far more complicated to analyze than those that add to the control-input ("matched" uncertainties). Therefore, it is common in control development to assume that $\hat{B}(x) \equiv B(x)$, yielding a closed-loop error dynamic of,

$$\dot{e} = \dot{r} - f(x) - B(x)B^+(x)\Big(Ke + \dot{r} - \hat{f}(x)\Big)$$
$$= -Ke + \hat{f}(x) - f(x)$$

Examples of application-specific adaptive control designs that ignore actuator dynamic uncertainties are [7] and [15]. Some results like [16] do handle unmatched uncertainties, but since that is not the focus of this thesis, we will not discuss their complexities here.

It should be noted that our restriction to control-affine dynamics was only to make more clear what a feedback linearization solution can look like. The logic above also applies to a variety of non-control-affine systems, provided they are feedback linearizable ($\exists u \mid \dot{e} = -Ke$). In general, we are just solving a dynamic inversion problem to devise a control law that results in the desired error convergence, with the adaptive update handling the solution's severe dependence on the model estimates. For the sake of clarity, we will only address control-affine dynamics in this thesis.

# CHAPTER 3
# UNCERTAINTY MODELING

Before devising an adaptive update for $\hat{f}$, we must discuss how uncertain model functions are even represented so that they can be manipulated. In traditional adaptive control, there two types of uncertainties: structured and unstructured.

Speaking abstractly, a structured uncertainty is any unknown function $a(x)$ that can be exactly represented by a finite linear combination of known, linearly-independent functions.

$$a(x) = \sum_{i=1}^{l} y_i(x)\theta_i$$

The set of known functions $\{y_1, y_2, \ldots, y_l\}$ is called the basis, while the unknown scalars $\theta_1, \theta_2, \ldots, \theta_l$ are called the weights. A more compact notation is,

$$a(x) = Y(x)\theta$$

where

$$Y(x) = \begin{bmatrix} y_1(x) & y_2(x) & \ldots & y_l(x) \end{bmatrix}$$

is called the regression matrix and

$$\theta = \begin{bmatrix} \theta_1 & \theta_2 & \cdots & \theta_l \end{bmatrix}^T$$

is a vector of the unknown weights. Physically speaking, structured uncertainty weights correspond to any uncertain parameters that the system is linear in. This means that the basis for a structured uncertainty is explicitly derived from physics. It is critical that the basis functions are linearly independent; breaking this condition is overparameterization and can hurt performance [2].

If the basis functions are also unknown, then $a(x)$ is called unstructured. Unstructured uncertainties correspond to terms that are nonlinear in any uncertain parameters, or terms that are completely unknown like external disturbances. The entire representation of an unstructured uncertainty falls under the art-form that is function approximation, making use of anything from Galerkin methods to neural networks. For example, an unstructured uncertainty may be *approximately* represented by a shallow multilayer perceptron,

$$a(x) \approx \Theta_2 \sigma(\Theta_1 x)$$

where $\Theta_1$ and $\Theta_2$ are matrices constructed from the elements of $\theta$, and $\sigma$ is an elementwise activation function.

It may seem odd that when a system has nonlinearities in any uncertain parameters, those terms are treated with very general function approximation tools. There is still some known structure to take advantage of, it is just not linear structure. Some adaptive control designs handle nonlinear-parametric uncertainty without arbitrary function approximation [17], but many do not distinguish it because taking special care of a specific type of parametric nonlinearity limits the generality of the result. It is simpler to umbrella them as "unstructured" and only derive an adaptive update for one type of nonlinear function approximator.

In observer-side parameter estimation, there is not as much talk of structured vs. unstructured uncertainty. Instead, focus is on the observability of $\theta$, which depends on the sensors available. If noise-free sensor outputs are sufficient to determine $\theta$ in finite-time, then $\theta$ is deemed observable. For linear systems, a mathematical condition for observability is easily derived [18]. For nonlinear systems, the mathematical formulation is more complicated but nonetheless possible and well-studied [19]. Regardless, we will explain later how if some perfect "state sensor" were available (which many adaptive control results imply by assuming exact state knowledge), then all structured uncertainty weights that influence the state trajectory are observable, and can even be exponentially estimated by a linear observer.

In any case, a finite vector of weight estimates, denoted as $\hat{\theta}$, is sufficient to represent our uncertain model functions in the adaptive update. That is, the adaptive update will manipulate $\hat{\theta}$ by selection of $\dot{\hat{\theta}}$, but will not change any basis functions. Since the true (or ideal) $\theta$ does not vary with time, the adaptive update is always a parameter estimator, whether the "parameters" are weights in a neural network or physical quantities like drag coefficients.

# CHAPTER 4
# CONTROLLER-SIDE PARAMETER ESTIMATION

In most control systems, an observer is used to make $\hat{x}$ track $x$. If this tracking is perfect (or "good enough"), then we can ignore the observer in the development of the adaptive update. We will refer to an adaptive update decoupled from the observer as controller-side parameter estimation.

In this section we will first examine one of the most basic adaptive updates for controller-side parameter estimation: tracking error gradient descent. Then we will examine an advancement of it known as concurrent learning. Both of these methods are motivated by Lyapunov analysis.

Lyapunov analysis is a technique for determining the stability of a dynamical quantity, like our tracking error $e(t)$. We will be concerned with proving one of three fundamental types of stability, listed here in order of increasing restrictiveness (and desirability):

1. Lyapunov stable: $\forall \epsilon > 0,\ \exists \delta > 0 \mid ||e(0)|| < \delta \implies ||e(0)|| < \epsilon\ \forall t \geq 0$

2. Asymptotically stable: Lyapunov stable and $||e(0)|| < \delta \implies \lim\limits_{t \to \infty} e(t) = 0$

3. Exponentially stable: asymptotically stable and
$\exists \{\alpha > 0, \beta > 0\} \mid ||e(0)|| < \delta \implies e(t) \leq \alpha e(0) \exp(-\beta t)\ \forall t \geq 0$

The stability condition is considered global if it holds for any $e(0)$. It can be shown that *if* there exists a scalar function $V(e) : \mathbb{R}^n \to \mathbb{R}$ such that,

- $V(e) = 0 \iff e = 0$
- $V(e) > 0 \iff e \neq 0$
- $\dot{V}(e) \leq 0$

then $e(t)$ is Lyapunov stable [20], and $V(e)$ is called a Lyapunov function. Furthermore, if $\dot{V}(e) < 0\ \forall e \neq 0$ *or* if $e(t)$ is uniformly continuous and $L_2$-integrable, then $e(t)$ is asymptotically stable [21]. Finally, if $||e|| \to \infty \implies |V(e)| \to \infty$ (a condition called radial-unboundedness) then the stability is global. All of these notions are made intuitive if $V$ is thought of as proportional to the "energy" of the system.

## 4.1 Tracking Error Gradient Descent

Tracking error gradient descent (TEGD) is a generalized view of the classic model-reference adaptive control (MRAC) architecture [22]. The core idea is to incrementally perturb $\hat{\theta}$ in the direction that results in the largest reduction of a tracking error quadratic, $\frac{1}{2}e^T e$.

We will first consider the case where $f$ is a completely structured uncertainty, and then we will consider the case where $f$ is a completely unstructured uncertainty. However, the two results can be trivially combined to handle an $f$ that consists of both structured terms and unstructured terms.

### 4.1.1 Structured Uncertainty

If the uncertainty $f$ is structured,

$$\hat{f} = Y(x)\hat{\theta}$$

then the closed-loop error dynamic can be written as,

$$\dot{e} = -Ke + Y(x)\tilde{\theta}$$

where $\tilde{\theta} = \theta - \hat{\theta}$ is the model error. Consider the candidate Lyapunov function,

$$V(e, \tilde{\theta}) := \frac{1}{2}e^T e + \frac{1}{2}\tilde{\theta}^T \Gamma^{-1} \tilde{\theta}$$

where $\Gamma > 0$ is a tuned learning rate matrix. We select this function because we are trying to prove that $e(t)$ is stable while simultaneously taking into account the behavior of our model error, in hopes of deriving a useful expression for $\dot{\hat{\theta}}$. The time derivative is,

$$\dot{V} = e^T \dot{e} + \tilde{\theta}^T \Gamma^{-1} \dot{\tilde{\theta}}$$
$$= e^T(-Ke + Y(x)\tilde{\theta}) - \tilde{\theta}^T \Gamma^{-1} \dot{\hat{\theta}}$$

where we made use of the fact that $\dot{\tilde{\theta}} = \dot{\theta} - \dot{\hat{\theta}} = -\dot{\hat{\theta}}$. If we design our adaptive update as,

$$\dot{\hat{\theta}} := \Gamma Y^T(x) e$$

then the Lyapunov derivative becomes,

$$\dot{V} = -e^T K e + e^T Y(x)\tilde{\theta} - \tilde{\theta}^T \Gamma^{-1} \Gamma Y^T(x) e$$
$$= -e^T K e \leq 0 \quad \forall \{e \neq 0, \tilde{\theta} \neq 0\}$$

$V$ is a function of both $e$ and $\tilde{\theta}$, but if $e = 0$ then $\dot{V} = 0$ regardless of the value of $\tilde{\theta}$. This means that $\dot{V}$ is negative semidefinite in $e$ and $\tilde{\theta}$, so we can only conclude that they are globally Lyapunov stable. Fortunately, $e(t)$ and $\tilde{\theta}$ being Lyapunov stable reveals that $V(e,\tilde{\theta})$ is bounded, so we have

$$\int e^T K e \, dt = -\int \dot{V} dt = V(0) - V(e,\tilde{\theta}) \in L_\infty \implies e(t) \in L_2$$

Therefore, $e(t)$ is globally asymptotically stable. This is a good statement about the robustness of the design (within our assumptions about the uncertainties), but it says nothing about performance. There are still the gain matrices $K$ and $\Gamma$ to be chosen by manual tuning on a specific system.

Moreover, we were unable to prove asymptotic stability of $\tilde{\theta}$, meaning that there is no guarantee of accurate system identification. After all, this adaptive update law just uses $Y^T(x)$ to map the $||e||$ descent direction to $\hat{\theta}$-space. There is no pressure on the controller to obtain an accurate estimate of $\theta$; there is only pressure to eliminate tracking error. For this design, the additional assumption we need in order to show asymptotic stability of $\tilde{\theta}$ is given in [2] as persistence of excitation,

$$\exists \Delta t > 0 \ \Big| \ \int_t^{t+\Delta t} Y^T(x) Y(x) d\tau > 0, \quad \forall t \geq 0$$

This is a dependency on how modally "exciting" the state trajectory is, *forever*. A more desirable, practical condition would be that the state trajectory be sufficiently exciting over one $\Delta t$ rather than every $\Delta t$.

### 4.1.2 Unstructured Uncertainty

If the uncertainty $f$ is unstructured, then we must first pick a suitable form for its parameterization. "Suitable" is a very situation-specific condition, so we will just present one common choice: a neural network. (Another common choice is a linear combination of radial basis functions). Thanks to the Universal Approximation Theorem, a three-layer, feedforward neural network with bias can approximate any compact function arbitrarily closely [23]. Thus we choose,

$$\hat{f}(x) := \hat{\Theta}_2 \sigma(\hat{\Theta}_1 \underline{x})$$

where $\underline{x} = \begin{bmatrix} x & 1 \end{bmatrix}^T$ is the bias-augmented state vector, $\sigma(v)$ is a differentiable elementwise activation function such as $\begin{bmatrix} \tanh(v_1) & \tanh(v_2) & \ldots & 1 \end{bmatrix}^T$, and $\hat{\Theta}_1$ and $\hat{\Theta}_2$ are matrices constructed from the elements of $\hat{\theta}$.

Note that we can't expect exact cancellation of $f(x)$ with $\hat{f}(x)$ because there is no guarantee that our neural network architecture admits a choice of weights to achieve perfect approximation. Thus, the best result we can hope for is Lyapunov stability of $e$ and $\tilde{\theta}$. The benefit is that we barely have to know about the nature of $f$ ahead of time (in contrast to the structured uncertainty case where we had to know or derive $Y$ from $f$).

Our parameter estimates are now packed into matrices, so it is convenient to consider the candidate Lyapunov function,

$$V(e, \tilde{\theta}) := \frac{1}{2} e^T e + \frac{1}{2} \text{tr}\left(\tilde{\Theta}_1^T \Gamma_1^{-1} \tilde{\Theta}_1\right) + \frac{1}{2} \text{tr}\left(\tilde{\Theta}_2^T \Gamma_2^{-1} \tilde{\Theta}_2\right)$$

where $\text{tr}(\cdot)$ is the trace function. The Lyapunov derivative is then,

$$\dot{V} = -e^T K e + e^T f(x) - e^T \hat{\Theta}_2 \sigma(\hat{\Theta}_1 \underline{x}) - \text{tr}\left(\tilde{\Theta}_1^T \Gamma_1^{-1} \dot{\hat{\Theta}}_1\right) - \text{tr}\left(\tilde{\Theta}_2^T \Gamma_2^{-1} \dot{\hat{\Theta}}_2\right)$$

The rest of the development details are given in [24], but the resulting adaptive update is,

$$\dot{\hat{\Theta}}_2 = \left(\Gamma_2 \sigma(\hat{\Theta}_1 \underline{x}) e^T\right)^T$$
$$\dot{\hat{\Theta}}_1 = \left(\Gamma_1 \underline{x} e^T \hat{\Theta}_2 \sigma'(\hat{\Theta}_1 \underline{x})\right)^T$$

where $\sigma'(v) = \frac{d\sigma}{dv}$. Euler integration of this adaptive update law is actually the infamous backpropagation algorithm commonly used by the machine learning community for training neural networks. Here, the tracking error is being backpropagated. Unfortunately, the only guarantee from this adaptive update is Lyapunov stability of the tracking error.

### 4.2 Concurrent Learning

The above TEGD methods are the foundation for dozens of other more complicated schemes. One of these is concurrent learning, which was first introduced in [3]. The idea is to add on a term to the update that depends on a "history stack" of recorded system data. The integration of this term serves as an iterative linear regression on model prediction error. To derive it, we have to look back at the original system dynamics,

$$\dot{x} = f(x) + B(x)u$$

If $f(x)$ is a structured uncertainty, then we can find another regression matrix $Y_u(x, \dot{x})$ such that,

$$Y_u(x, \dot{x})\theta = B(x)u$$

Note that computing $Y_u(x, \dot{x})$ requires knowing both the state and the state derivative. In some cases, sensors like accelerometers make state derivative measurements available, but in other cases a numerical approximation must be used. (There is actually a slightly more complicated method known as integral concurrent learning that provides a less ad-hoc solution [25]). For simplicity, we will assume that the state derivative is known and define the concurrent learning adaptive update as,

$$\dot{\hat{\theta}} := \Gamma Y^T(x)e + \Gamma_u \sum_{i=1}^{N} Y_u^T(x_i, \dot{x}_i)\Big(B(x_i)u_i - Y_u(x_i, \dot{x}_i)\hat{\theta}\Big)$$

where the data set $\{(x_1, \dot{x}_1, u_1), (x_2, \dot{x}_2, u_2), \ldots, (x_N, \dot{x}_N, u_N)\}$ is the history stack and $\Gamma_u > 0$ is another learning gain matrix. Note that the left-hand term is just the TEGD update for structured uncertainty.

The difference inside the right-hand term is a known linear transformation of the prediction error, and the prediction error is exactly equal to the model error if all of our assumptions (including state knowledge) are correct,

$$B(x)u - Y_u(x, \dot{x})\hat{\theta} = Y_u(x, \dot{x})\tilde{\theta}$$

$$\implies \sum_{i=1}^{N} Y_u^T(x_i, \dot{x}_i)\Big(B(x_i)u_i - Y_u(x_i, \dot{x}_i)\hat{\theta}\Big) = \sum_{i=1}^{N} Y_u^T(x_i, \dot{x}_i)Y_u(x_i, \dot{x}_i)\tilde{\theta}$$

This makes it clear that *if*,

$$\sum_{i=1}^{N} Y_u^T(x_i, \dot{x}_i)Y_u(x_i, \dot{x}_i) > 0$$

then the update is just,

$$\dot{\hat{\theta}} = \Gamma Y^T(x)e + \Gamma'_u \tilde{\theta}$$

where $\Gamma'_u = \Gamma_u \sum_{i=1}^{N} Y_u^T(x_i, \dot{x}_i)Y_u(x_i, \dot{x}_i)$ is just another positive definite matrix. Intuitively, as tracking error goes to zero, the right-hand term dominates and the parameter estimates will head toward their true values, while if an unaccounted-for disturbance increases tracking error, stability will be retained by the robust TEGD left-hand term.

To actually prove these claims, we can use the same Lyapunov function we constructed for structured TEGD,

$$V(e, \tilde{\theta}) := \frac{1}{2}e^T e + \frac{1}{2}\tilde{\theta}^T \Gamma^{-1} \tilde{\theta}$$

Now the Lyapunov derivative is,

$$\dot{V} = e^T(-Ke + Y(x)\tilde{\theta}) - \tilde{\theta}^T \Gamma^{-1} \dot{\hat{\theta}}$$

$$= -e^T Ke + e^T Y(x)\tilde{\theta} - \tilde{\theta}^T \Gamma^{-1}\Big(\Gamma Y^T(x)e + \Gamma_u \sum_{i=1}^{N} Y_u^T(x_i, \dot{x}_i)Y_u(x_i, \dot{x}_i)\tilde{\theta}\Big)$$

$$= -e^T Ke - \tilde{\theta}^T \Big(\Gamma^{-1}\Gamma_u \sum_{i=1}^{N} Y_u^T(x_i, \dot{x}_i)Y_u(x_i, \dot{x}_i)\Big)\tilde{\theta}$$

Because both gain matrices are positive definite, $\Gamma^{-1}\Gamma_u$ must be positive definite. Thus, as long as $\sum_{i=1}^{N} Y_u^T(x_i, \dot{x}_i)Y_u(x_i, \dot{x}_i)$ is also positive definite, then the Lyapunov derivative will be negative definite, proving that both $e$ and $\tilde{\theta}$ are globally asymptotically stable. Further, it is shown in [3] that the convergence can be upper-bounded by an exponential (i.e. global exponential stability)!

So concurrent learning achieves exponential convergence of both tracking error and model error without requiring persistence of excitation. However, there is still a burning question: how do we choose the values in the history stack? Practical memory limitations prevent $N$ from being huge, so there needs to be a data selection process.

The answer may seem obvious at first. Since we want $\sum_{i=1}^{N} Y_u^T(x_i, \dot{x}_i)Y_u(x_i, \dot{x}_i) > 0$, we should just store data that raises the minimum eigenvalue of that matrix sum– data which may be considered rich in new information.

This is ideal in theory, but problematic in practice. Concurrent learning relies on a number of assumptions, from perfect state estimation to the linear structure of the uncertain parameters to the accuracy of the actuator model. When these assumptions are not met in practice, a $Y_u^T(x_i, \dot{x}_i)Y_u(x_i, \dot{x}_i)$ value can appear rich in information while it is really just erroneous data.

There is a decent amount of research in this area. For example, [26] provides a data selection algorithm that considers state estimation error. Unfortunately, it only considers that the full state estimate gets better with time to motivate a greedy purging algorithm, rather than using any probabilistic information about the individual states. Regardless, all ideas so far are in some sense heuristic.

# CHAPTER 5
# STOCHASTIC OBSERVERS

Continuous-time stochastic analysis requires some very advanced math like Itô calculus. To make this document more accessible, we will handle our discussion of stochastic observers in discrete-time, but understand that this is barely a restriction. Modern control systems are almost always implemented on digital electronics, so even the continuous-time adaptive update laws previously discussed will need to be discretized for integration at some point. More than anything, the switch to discrete-time here is just a change in the order of operations: discretizing and then analyzing instead of analyzing and then discretizing.

Consider the following hidden-state, discrete-time stochastic system with process $\phi$ and sensor $h$,

$$x_{t+1} = \phi(x_t, u_t, t, \eta_t)$$

$$z_t = h(x_t, u_t, t, \nu_t)$$

where $\eta_t$ is process noise, $\nu_t$ is sensor noise, and $x_0$ is our $t=0$ prior for the state vector $x$. The known probability distributions of these assumed *independent* random variables encode our uncertainty in / partial knowledge of the process model, sensor model, and initial condition respectively. The only variables known with perfect certainty are the sensor measurements $z_t$ we actually received, the control-input $u_t$ we chose, and the current time $t$.

Letting $\rho(a|b)$ denote the probability distribution of random variable $a$ conditioned on events $b$, the Markovian nature of the process model asserts that

$$\rho(x_{t+1}|x_t, x_{t-1}, \ldots, x_0, u_t, u_{t-1}, \ldots, u_0) = \rho(x_{t+1}|x_t, u_t)$$

We refer to this quantity as a transition probability. Let $\mathscr{Z}_t$ denote the set $\{z_0, z_1, \ldots, z_t\}$ of all sensor measurements up to time $t$. Then the probability distribution for the next state given only the measurements up to now is,

$$\rho(x_{t+1}|\mathscr{Z}_t, u_t) = \int_{\mathbb{R}^n} \rho(x_{t+1}|x_t, u_t)\rho(x_t|\mathscr{Z}_t)dx_t$$

This equation is called the predict step because it computes the probability distribution of the state one step into the future beyond the latest measurement. It is a product of the probability distribution for the current state (given the measurements we know so far) and the transition probability, summed over all possible current states. I.e. it is just a direct computation of unions and intersections on the underlying probability space.

When a new measurement for time $t+1$ comes in, we can use Bayes' rule [8] to show that,

$$\rho(x_{t+1}|\mathscr{Z}_{t+1}) = \frac{\rho(z_{t+1}|x_{t+1})\rho(x_{t+1}|\mathscr{Z}_t, u_t)}{\rho(z_{t+1}|\mathscr{Z}_t)}$$

where the denominator is just the integral of the numerator with respect to $x_{t+1}$. This equation is called the correct step because we are correcting our prediction with the new measurement information. The distribution $\rho(z_{t+1}|x_{t+1})$ is called the likelihood of the measurement $z_{t+1}$ and it can be computed from the function $h$.

The predict and correct equations together form an algorithm called recursive Bayesian estimation (RBE), and it is the core of stochastic observers. RBE is not a state estimation method itself. Rather, it is a method for evolving the probability distribution $\rho(x_t|\mathscr{Z}_t)$ as new measurements are received. In other words, RBE is the dynamic for $\rho(x_t|\mathscr{Z}_t)$, which is often called the belief distribution. The importance of RBE is in the fact that the belief distribution is a sufficient statistic for $x$; it contains all the information needed to compute any type of probabilistic $x$-estimate [8].

The arguably "best" estimate of $x$ is the maximum-likelihood estimate (the belief mode),

$$\hat{x}_t^* = \arg\max_{x_t}\Big(\rho(x_t|\mathscr{Z}_t)\Big)$$

Another option is the expected value (the belief mean),

$$\hat{x}_t = \mathrm{E}(x_t|\mathscr{Z}_t) = \int_{\mathbb{R}^n} x_t \rho(x_t|\mathscr{Z}_t) dx_t$$

The maximum-likelihood estimate is technically preferred because if the belief distribution is highly multimodal, then the expected value may not actually be very likely. However, for symmetric, unimodal distributions like the Gaussian, these estimates are identical, and unimodal belief distributions with at least crude symmetry are not uncommon.

Unfortunately, RBE is a very challenging set of equations to implement for continuous state spaces like $\mathbb{R}^n$. The infinite multidimensional integrals make most numerical methods intractable. The common approach then, is to select assumptions about the system that make analytical simplifications of RBE possible.

## 5.1 The Kalman Filter

One nice simplification would be if the belief distribution always had some fixed parametric form. For example, if the belief distribution was always Gaussian (parameterized by a mean and covariance), then the RBE integrals could be easily computed. Let,

$$\hat{x}_t := \mathrm{E}(x_t|\mathscr{Z}_t) = \int_{\mathbb{R}^n} x_t \rho(x_t|\mathscr{Z}_t)dx_t$$

denote the expected value (mean) of $x_t$ given everything we know so far and let,

$$P_t := \mathrm{cov}(x_t, x_t|\mathscr{Z}_t) = \mathrm{E}\left((x_t - \hat{x}_t)(x_t - \hat{x}_t)^T|\mathscr{Z}_t\right) = E(x_t x_t^T|\mathscr{Z}_t) - \hat{x}_t \hat{x}_t^T$$

denote the covariance of $x_t$, i.e. the second central-moment of the belief distribution. For any linear transformation $A$, the linearity of the expected value integral imposes that,

$$\mathrm{E}(Ax_t|\mathscr{Z}_t) = A\hat{x}_t$$

$$\begin{aligned}
\mathrm{cov}(Ax_t, Ax_t|\mathscr{Z}_t) &= E\left((Ax_t)(Ax_t)^T|\mathscr{Z}_t\right) - A\hat{x}_t(A\hat{x}_t)^T \\
&= AE\left(x_t x_t^T|\mathscr{Z}_t\right)A^T - A\hat{x}_t \hat{x}_t^T A^T \\
&= AP_t A^T
\end{aligned}$$

It can also be shown that linear combinations of Gaussian random variables are still Gaussian [8] (a fact fundamentally related to the Central Limit Theorem). Therefore if,

- $x_t$ starts Gaussian: $\rho(x_0) = \mathcal{N}(\hat{x}_0, P_0)$
- $\eta_t$ and $\nu_t$ are Gaussian: $\rho(\eta_t) = \mathcal{N}(\bar{\eta}_t, Q_t)$ and $\rho(\nu_t) = \mathcal{N}(\bar{\nu}_t, R_t)$
- The dynamics are linear in all random variables:

$$x_{t+1} = \Phi_t x_t + \phi_u(u_t, t) + G_t \eta_t$$

$$z_t = H_t x_t + h_u(u_t, t) + L_t \nu_t$$

then the belief distribution is Gaussian for all time, and thus always parameterizable by just its mean $\hat{x}_t$ and covariance $P_t$. This is called the linear-Gaussian assumption. Now, using our equations for the means and covariances of linearly transformed random variables, the RBE predict step can be reduced to,

$$\hat{x}_{t+1}^- := \mathrm{E}(x_{t+1} | \mathscr{Z}_t, u_t)$$
$$= \Phi_t \hat{x}_t + \phi_u(u_t, t) + G_t \bar{\eta}_t$$

$$P_{t+1}^- := \mathrm{cov}(x_{t+1}, x_{t+1} | \mathscr{Z}_t, u_t)$$
$$= \Phi_t P_t \Phi_t^T + G_t Q_t G_t^T$$

and the RBE correct step can be reduced to,

$$\hat{x}_{t+1} = \mathrm{E}(x_{t+1} | \mathscr{Z}_{t+1})$$
$$= \hat{x}_{t+1}^- + \mathsf{K}_{t+1}\left(z_{t+1} - \hat{z}_{t+1}\right)$$

$$P_{t+1} = \mathrm{cov}(x_{t+1}, x_{t+1} | \mathscr{Z}_{t+1})$$
$$= (I - \mathsf{K}_{t+1} H_{t+1}) P_{t+1}^-$$

where,

$$\mathsf{K}_t := P_t^- H_t^T (H_t P_t^- H_t^T + L_t R_t L_t^T)^{-1}$$

$$\hat{z}_t := H_t \hat{x}_t^- + h_u(u_t, t) + L_t \bar{\nu}_t$$

(These equations often appear in a less bulky form where it is assumed that the noises are zero-mean and purely additive: $\bar{\eta}_t = 0$, $\bar{\nu}_t = 0$, $G_t = I$, and $L_t = I$). With the belief distribution $\mathcal{N}(\hat{x}_t, P_t)$ computable at all times, we can pick our state estimate. The belief distribution is Gaussian, so the mean and mode coincide, leaving $\hat{x}_t$ itself as the obvious choice.

This recursion is the Kalman filter. It is a linear observer with a time-varying gain $K_t$ called the Kalman gain. Put simply, the Kalman filter uses the covariances of the process noise, sensor noise, and current state estimate to best distinguish the cause of sensor prediction error $(z - \hat{z})$ and update the state estimate accordingly. Typically, the sensor noise covariance is computed offline by testing the sensor, while the process noise covariance acts as a "tuning knob" for our confidence in the model's accuracy. However, both could be computed with online statistics.

The word "best" was used here because under the linear-Gaussian assumption, the Kalman filter yields the minimum mean squared-estimation-error (MSEE). In other words, the estimate it produces has the smallest possible covariance ellipsoid. This is no surprise, as here we have seen the Kalman filter as the linear-Gaussian case of RBE; it computes the mean of the Gaussian belief distribution, which coincides with the maximum-likelihood estimate.

Interestingly, the Kalman filter's optimality can also be proven by applying the orthogonality principle to the Hilbert space of random variables, which is actually how it was first derived [27]. In fact, that proof shows that *of all linear observers*, the Kalman filter is the global minimizer of MSEE regardless of whether the belief distribution is Gaussian. However, in the non-Gaussian case, even the best *linear* observer may be a very bad one, so we are inclined to look at nonlinear options too. Moreover, the Kalman filter is really only defined for linear systems, so for nonlinear systems (Gaussian or not), modifications are necessary.

### 5.2 Nonlinear Extensions

The simplest way to extend the Kalman filter to nonlinear systems is to iteratively apply the Kalman filter equations to a local linearization of the system dynamics. Specifically, we linearize about the current state estimate by letting,

$$\Phi_t = \frac{\partial \phi}{\partial x}\bigg|_{\hat{x}_t, u_t, t, \bar{\eta}_t} \qquad G_t = \frac{\partial \phi}{\partial \eta}\bigg|_{\hat{x}_t, u_t, t, \bar{\eta}_t}$$
$$H_t = \frac{\partial h}{\partial x}\bigg|_{\hat{x}_t, u_t, t, \bar{\nu}_t} \qquad L_t = \frac{\partial h}{\partial \nu}\bigg|_{\hat{x}_t, u_t, t, \bar{\nu}_t}$$

The resulting algorithm is called the extended Kalman filter (EKF). For many single-body systems (boats, submarines, satellites, quadcopters, small ground vehicles, etc), the primary nonlinearity due to orientation lends itself well to iterative linearization, so the EKF has become ubiquitous in practice. Furthermore, the above Jacobian matrices are often sparse, which allows for great computational simplifications [28].

For systems with trickier nonlinearities, the unscented Kalman filter (UKF) is a commonly used alternative. Notice that the EKF first approximates the dynamics and then applies the exact transformation law for the belief mean and covariance. The idea behind the UKF is to use the exact nonlinear dynamics and instead apply an approximate transformation law for the belief mean and covariance. This has been shown to outperform the EKF in a number of situations [29].

What makes this idea possible is the "unscented transform," which approximately computes the mean and covariance of a random variable after an arbitrary nonlinear transformation. It does this by first representing the mean and covariance of the untransformed random variable as a deterministic set of "sigma points" (not to be confused with Monte Carlo sample points which are obtained randomly). The sigma points are then propagated through the nonlinear transformation and their mean and covariance recomputed. This is shown schematically in Figure 5-1 (at the end of this section). For a more detailed review of the UKF, see [29].

Ultimately, anything based on the Kalman filter is just an estimator of the belief mean and covariance. Looking at Figure 5-1, we see that even though the unscented transform accurately computed the next mean, the mean itself is not in a region of high belief density (made clear by the sampled distribution on the left). The mean can be a poor state estimate if the true belief distribution is multimodal or highly asymmetric. For these cases, it is common to use a particle filter (also called the sequential Monte Carlo method).

The particle filter relies on being able to randomly sample a very large set of states from the belief distribution, and then process / update them all to estimate the next belief distribution, from which we can deduce the next maximum likelihood estimate. If the number of samples is large enough, the results can be extremely accurate with barely any assumptions.

Unfortunately, particle filters are very computationally expensive (the number of samples needed grows exponentially with the state dimension), and so for systems with a large number of states, they can be intractable for realtime work.

While it is possible to derive a more system-tailored nonlinear stochastic observer from Lyapunov analysis, it can be very difficult and end up relying on its own slew of assumptions, so most designers just work within the limits of the stability proofs for the EKF, UKF, and particle filter (see [12]).

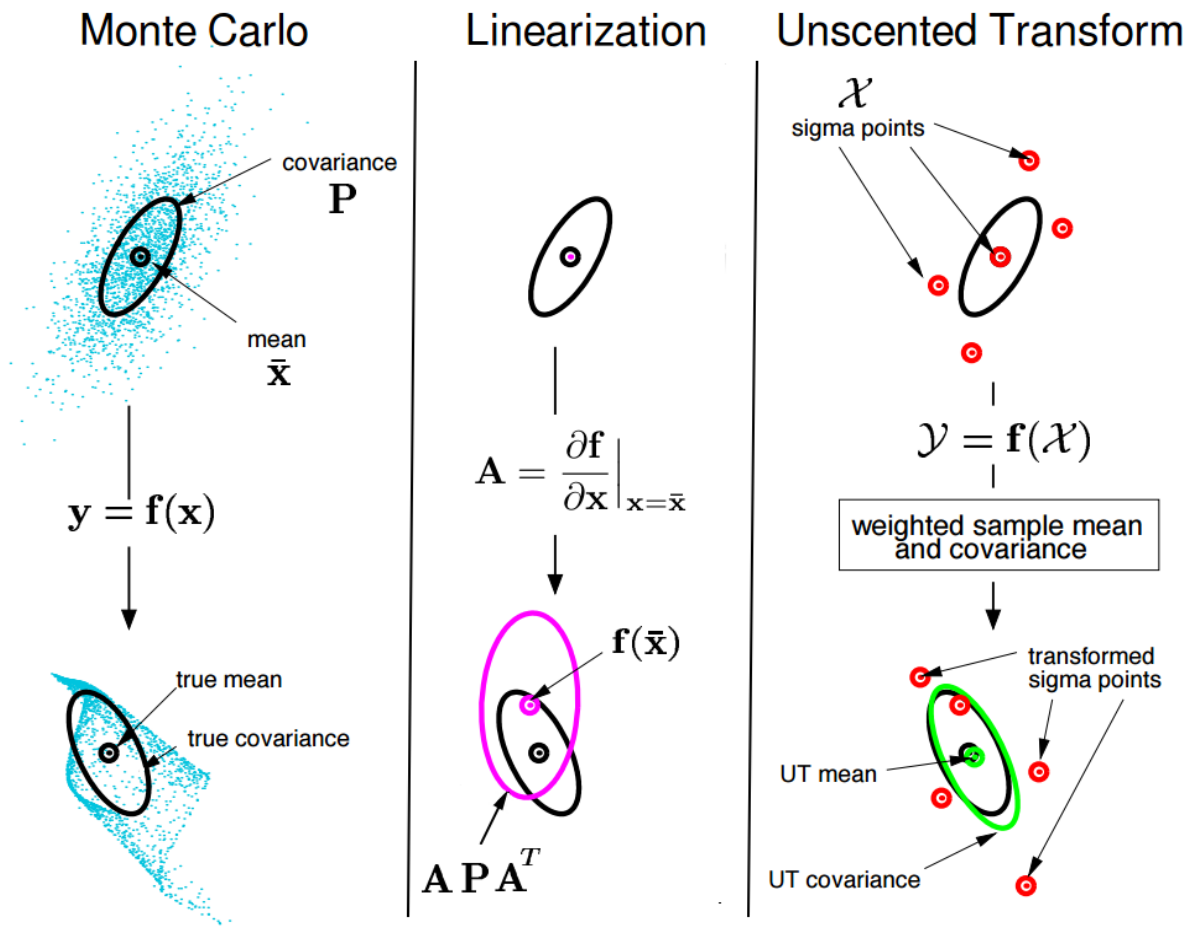

Figure 5-1. A function $f$ is applied to a random variable $x$ with mean $\bar{x}$ and covariance $P$. Monte Carlo (left) randomly samples a ton of points (blue dots) from the original distribution and passes them all through $f$ to form a sample-distribution from which the $f$-transformed mean and covariance can be recomputed with high accuracy. Linearization (middle) approximates $f$ by its Jacobian $A$ and computes the transformed mean and covariance analytically. The unscented transform (right) computes a deterministic set of "sigma points" that capture the original mean and covariance statistics exactly, and then propagates them through $f$ to compute the transformed mean and covariance. The unscented transform is still approximate because the sigma points do not capture all of the higher moments of the original distribution. This image was adapted from [29].

# CHAPTER 6
# OBSERVER-SIDE PARAMETER ESTIMATION

The previously discussed stochastic observers are not limited to just state estimation. In their development, we simply sought to estimate some quantity $x$ given partial information about it. That partial information came in two forms: a differential equation that governs the evolution of $x$ and an algebraic function of $x$ who's output is known at various times. This information is partial because both equations are corrupted by noise– random variables that encode the nature of our uncertainty. (Additionally, the sensor function may not be invertible).

From that perspective, a vector of unknown parameters $\theta$ is just another example of an $x$; specifically one with a stationary process,

$$\theta_{t+1} = \theta_t$$

We can even throw in a noise term if we aren't completely sure that $\theta$ is truly constant. Either way, we don't have to switch our problem from state estimation to parameter estimation; we can handle both with the same framework. Define the "augmented-state" vector as,

$$\underline{x}_t := \begin{bmatrix} x_t \\ \theta_t \end{bmatrix}$$

Its dynamics can be expressed as,

$$\underline{x}_{t+1} = \underline{\phi}(\underline{x}_t, u_t, t, \eta_t) = \begin{bmatrix} \phi(x_t, u_t, t, \eta_t, \theta_t) \\ \theta_t \end{bmatrix}$$

$$z_t = \underline{h}(\underline{x}_t, u_t, t, \nu_t)$$

where the extra argument to $\phi$ indicates explicit use of the parameters $\theta_t$, and the new sensor function $\underline{h}$ may also make use of the parameter part of $\underline{x}_t$ (for example, sensor biases can be included in $\theta_t$). From here, stochastic observer development is no different than before. Furthermore, as the $\hat{\theta}$ part of the $\hat{\underline{x}}$ estimate converges, the predict step will become increasingly accurate, boosting the overall observer performance. I.e. not only will state estimation help parameter estimation, but parameter estimation will help state estimation. This is the tight sharing of information we motivated in the Introduction.

With our observer generating $\hat{\theta}_t$, we are ready to implement an adaptive controller simply by using those parameter estimates in a control law. But can we be guaranteed stability? Well, we were already planning to use the observer's $\hat{x}$ estimates in the controller anyway– were we guaranteed stability at that point? Few adaptive control results consider that the cancellation of $\hat{f}(\hat{x}) - f(x)$ is just as dependent on the accuracy of $\hat{x}$ as it is on the accuracy of $\hat{f}$. This concern discrepancy is mostly tied to experience: excellent state estimators have become so common that it is no-longer outlandish to assume $\hat{x} \equiv x$.

Of course, that rationale has no theoretical rigor and may someday cause a disaster. Unfortunately, it might not even be possible to derive a stability proof for the extremely general nonlinear stochastic system described above. We must make at least some assumptions or analyze a more specific system.

The previously derived controller-side parameter estimation techniques seemingly applied to a very general class of systems, but in actuality they relied on the strong assumption of exact state knowledge. Making that assumption on the observer-side would ridiculously oversimplify the problem. If $x$ is exactly known and the uncertainties are structured, then the estimation problem is just linear regression– a setting for Kalman filter optimality. I.e., the typical adaptive control assumptions guarantee that the basic Kalman filter converges $\tilde{\theta}$ faster than any other adaptive update law. For the unstructured case, we can look to [30] for an in-depth review of the EKF and UKF's success in training neural networks.

Regardless, such assumptions cannot really be made on the observer-side because no perfect state sensor exists. The observer will always have the job of estimating $x$; all we can do for adaptive control is have the observer estimate $\theta$ as well. This observer-side parameter estimation (sometimes called "dual estimation") is becoming increasingly popular in modern control systems [29]. We will now provide a simple simulation case-study with somewhat impressive results to demonstrate observer-side parameter estimation for adaptive control.

**Electric motor demonstration:** Consider a voltage-controlled electric motor for use on a robot arm or wheeled vehicle. Let $x_1$ be the angular position of the shaft and let $x_2$ be the angular velocity of the shaft. A simple dynamic for the state $x = [x_1 \ x_2]^T$ is,

$$\dot{x} = \begin{bmatrix} x_2 \\ m^{-1}(bu - cx_2 + d) \end{bmatrix}$$

where $m$ is the inertial load, $b$ is the voltage-to-torque transfer ratio, $u$ is the voltage we apply to the motor, $c$ is the rotational drag coefficient, and $d$ is some external disturbance torque. Our input voltage is limited by the power supply capabilities, $u \in [-u_{\text{lim}}, u_{\text{lim}}]$.

Performing a first-order discretization and adding noise to provide the effects of unconsidered forces, we have,

$$x(t + \Delta t) = x(t) + \begin{bmatrix} x_2(t) \\ m^{-1}\Big(bu(t) - cx_2(t) + d(t)\Big) \end{bmatrix} \Delta t + \begin{bmatrix} 0 \\ \omega_f(t) \end{bmatrix}$$

where $\omega_f \sim \mathcal{N}(0, q_f)$. To emulate the unknown nonlinearities of power transfer and friction, $b$ and $d$ are made time-varying through random-walk,

$$b(t + \Delta t) = b(t) + \omega_b(t)$$

$$c(t + \Delta t) = c(t) + \omega_c(t)$$

where $\omega_b(t) \sim \mathcal{N}(0, q_b)$ and $\omega_c(t) \sim \mathcal{N}(0, q_c)$. Two different $T$-long simulations will be run: one with $d(t)$ as a step disturbance and one with $d(t)$ as a sinusoidal disturbance.

The motor shaft is sensed by an optical encoder– a device that counts the number of interruptions ("ticks") of an optical signal as the motor spins a slotted disk. The encoder's resolution $a$ is the number of ticks per unit rotation, so the encoder tick-count $z$ can be expressed as,

$$z(t) = \text{floor}\Big(ax_1(t)\Big)$$

where floor($\cdot$) rounds its argument down to the nearest whole number. Technically, this encoder is an "absolute" encoder because it measures $x_1$ rather than an increment $\Delta x_1$.

The following numerical values were used for everything above. All units are base-SI (kg, s, m, N, V, etc...) and angles are in degrees.

$$T = 40, \quad \Delta t = 0.05$$

$$x(0) = [15 \ \ 0]^T, \quad u_{\text{lim}} = 50$$

$$m = 1, \quad a = {}^{512}/_{360}$$

$$b(0) = 2, \quad c(0) = 5$$

$$q_f = 10^{-4}, \quad q_b = 10^{-6}, \quad q_c = 10^{-3}$$

$$d_1(t) = 8\delta(t - 20), \quad d_2(t) = 3\cos(t + 2) + 3$$

Note that the sampling rate of $^1/_{\Delta t} = 20$ Hz and the encoder resolution of $512$ ticks per revolution are very realistic for an affordable motor-control system.

At any timestep, we will only be able to utilize exact knowledge of $u$, $z$, and $t$. However, we desire the performance benefits of not just full state feedback, but also exact model knowledge. Therefore, we define the augmented-state vector as,

$$\underline{x} = \begin{bmatrix} x_1 \\ x_2 \\ b/m \\ c/m \\ d/m \end{bmatrix}$$

and set out to estimate it with a stochastic observer. Note that we combined $m^{-1}$ with the other parameters to maintain observability. The dynamics are identical for any common scaling of $m$, $b$, $c$, and $d$. E.g., doubling all of them has no effect on the system. Therefore, $[m, b, c, d]^T$ would have been ill-defined for estimation (have unobservable components).

Having avoided that issue, our process model is,

$$\underline{x}(t+\Delta t) = \underline{\phi}\Big(\underline{x}(t), u(t), t, \eta(t)\Big) = \underline{x} + \begin{bmatrix} \underline{x}_2 \\ \underline{x}_3 u - \underline{x}_4 \underline{x}_2 + \underline{x}_5 \\ 0 \\ 0 \\ 0 \end{bmatrix} \Delta t + \eta$$

where $\eta \sim \mathcal{N}(0, Q)$. This models $b$, $c$, and $d$ as randomly-walking states, and while that works for $b$ and $c$, it is a poor representation of $d$. We are effectively fitting a constant to $d$ even though we know it can do almost anything. Fortunately, we can reflect this in our choice of $Q$ by placing a relatively high variance on our untrustworthy disturbance model. This is as opposed to, say, approximating the disturbance by a neural network who's weights are put in the augmented-state. We refrain from that here to specifically demonstrate what happens when a somewhat underparameterized model is used in the observer. We conservatively set our process noise variance to,

$$Q = \text{diag}(10^{-10},\ 10^{-3},\ 10^{-5},\ 10^{-2},\ 10^{-1})$$

We could also augment our state with the parameter $a$, but it would be highly unusual to not know the resolution of the encoder we bought, so we can rightfully assume that $a$ is known. This would incline us to make our sensor model identical to the simulation's sensor function. However, the floor($\cdot$) operation in that function is highly discontinuous and can lead to numerical issues in the observer, so we will opt for a differentiable approximation of the encoder's behavior: the line that runs through the center of the actually generated "staircase" signal,

$$\underline{h}(\underline{x}, u, t, \nu) = a\underline{x}_1 + \nu$$

where $\nu \sim \mathcal{N}(0, R)$. Here, the additive Gaussian sensor noise is a crude representation of the sort of "discretization error" caused by the encoder's finite-resolution nature. We select a standard deviation of $\sqrt{R} = 1$ tick.

The nonlinearities here are rather mild, so both the EKF or UKF should work fine. We choose the UKF just because it is a bit more general purpose. We guess our initial condition very incorrectly as,

$$\underline{\hat{x}}(0) = \begin{bmatrix} -15 \\ 10 \\ 1 \\ 1 \\ 0 \end{bmatrix}$$

but reflect our uncertainty in the initial covariance,

$$\underline{P}(0) = \text{diag}(50,\ 40,\ 10,\ 10,\ 50)$$

We want the motor to track the following state trajectory,

$$r(t) = \begin{bmatrix} 15\sin(0.5t) \\ 7.5\cos(0.5t) \end{bmatrix}$$

Our feedback-linearization control law is,

$$u = 5(r_1 - \underline{\hat{x}}_1) + 5(r_2 - \underline{\hat{x}}_2) + \frac{1}{\underline{\hat{x}}_3}\left(\dot{r} + \underline{\hat{x}}_4 r_2 - \underline{\hat{x}}_5\right)$$

The factor of $1/\underline{\hat{x}}_3$ corresponds to the $\hat{B}^+$ we had back when discussing general feedback linearization. It does raise a concern: what if $\underline{\hat{x}}_3 = 0$ even if just for an instant? For that to happen, the observer would need to think that either $b = 0$ (no control effectiveness) or $m = \infty$ (unmovable motor). We will see empirically that the observer has enough information to easily avoid those nonphysical estimates. However, as a precaution, our controller will ignore the feedforward term if the singularity occurs in a transient. Alternatively, we could do something like what is done in [16]: wait for the matched uncertainty estimates to converge and then handle the unmatched uncertainties.

The results of the first simulation (using a step disturbance) are shown in Figure 6-1. The results of the second simulation (using a sinusoidal disturbance) are shown in Figure 6-2. An example of the encoder sensor measurements is shown in Figure 6-3 to exhibit what the only information available looked like.

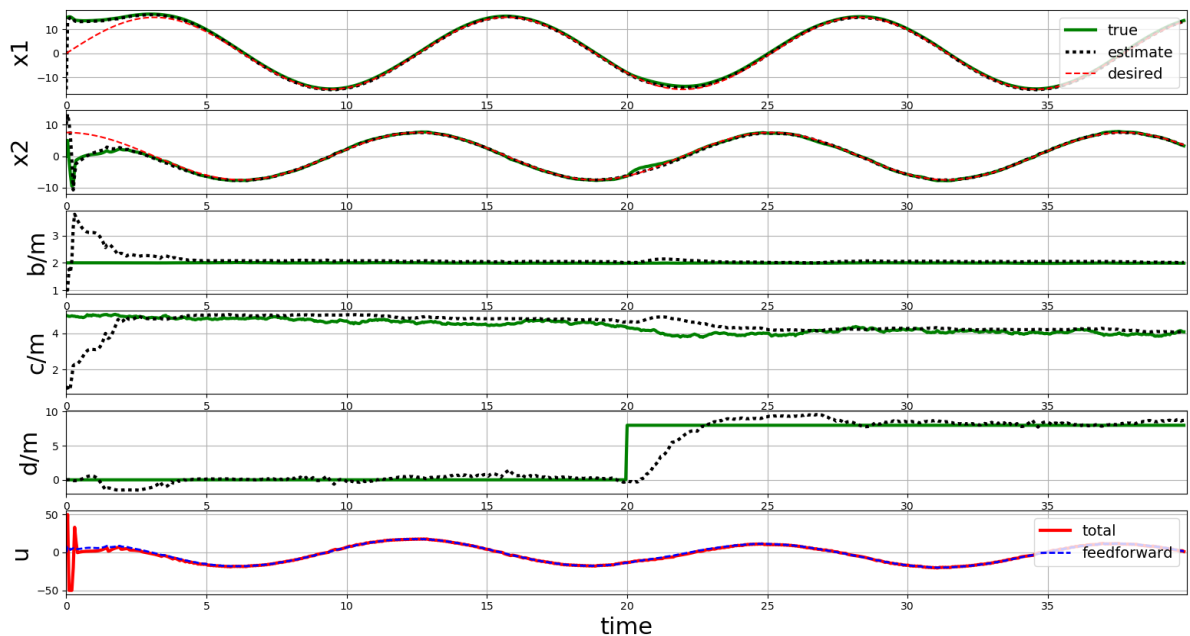

Figure 6-1. Plots of $\underline{x}$, $\underline{\hat{x}}$, $r$, and $u$. In the first five plots we see that $\underline{\hat{x}}$ (dotted-black) moved from its erroneous initial condition to $\underline{x}$ (green) very quickly. After the brief estimation transient, the feedback part of $u$ became negligible in favor of the accurate feedforward part, which caused the excellent trajectory tracking (green following dashed-red) seen in the first two plots. At $t = 20$ the true disturbance parameter $d/m$ jumped, but its change was estimated fine while barely disturbing the other estimates or trajectory tracking.

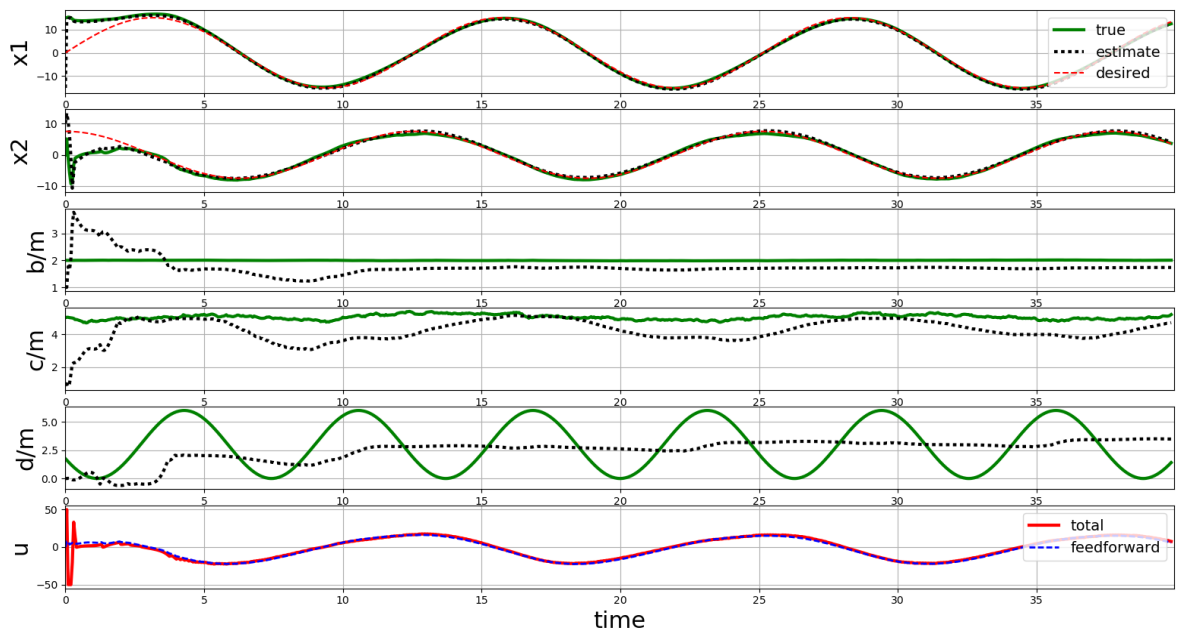

Figure 6-2. Results similar to Figure 6-1, but now the disturbance is a sinusoid the entire time. We only gave the UKF one degree-of-freedom to model $d/m$, so it makes sense that its estimate holds at roughly the average-value of the sinusoid. Regardless, the UKF is still fitting its model the best it can, and its fit is obviously adequate for control purposes: trajectory tracking is still excellent and the feedforward is still dominating $u$. I.e., even though the UKF's estimates of $b/m$, $c/m$, and $d/m$ are all "wrong" due to inadequate parametric freedom, their combined effect still yields an overall useful model. Heuristically speaking, it is the minimum MSEE *fit* at all times.

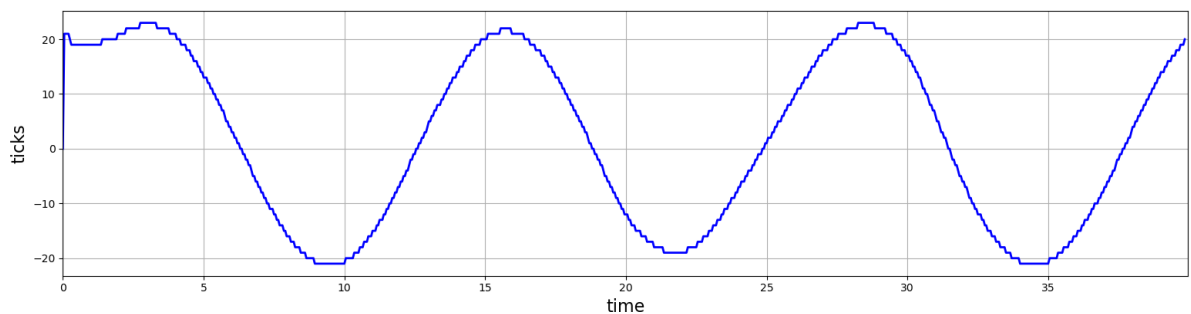

Figure 6-3. The encoder measurements during the simulation that generated Figure 6-1. These "stairstep" values were the *only* system measurements available, and yet the UKF was capable of producing excellent velocity estimates. (Consider that an ordinary finite difference method for velocity computation would completely fail on discontinuous measurements like these). The UKF's process model was critical to producing smooth state estimates, but at the same time, the UKF's parameter estimates were critical to the accuracy of its process model! By sharing all estimation information within the observer, the great results of Figures 6-1 and 6-2 were attainable.

# CHAPTER 7
# PROPOSITION: STOCHASTIC CONCURRENT LEARNING

So far, we have only provided heuristic and empirical support for the safety and effectiveness of observer-side parameter estimation for adaptive control. The main argument has been that controller-side adaptive updates hurt their estimates by decoupling them from probabilistic information, so the obvious solution is to put all estimation jobs in a single stochastic observer.

However, all stochastic observers lack a certain feature that most controller-side adaptive updates have: dependence on tracking error. Looking back at the concurrent learning adaptive update,

$$\dot{\hat{\theta}} := \Gamma Y^T(x)e + \Gamma_u \sum_{i=1}^{N} Y_u^T(x_i, \dot{x}_i)\Big(B(x_i)u_i - Y_u(x_i, \dot{x}_i)\hat{\theta}\Big)$$

we can associate the right-hand term with what an observer would do: regression. The left-hand term, motivated by Lyapunov analysis, is there to dominate parameter estimation if the tracking error starts growing. Under the assumptions of concurrent learning, another way to view this equation is,

$$\dot{\hat{\theta}} = \Gamma Y^T(x)e + \Gamma'_u(\theta - \hat{\theta})$$

Essentially, current learning says that if the history stack is sufficiently rich, then we already know our *best-fit* $\theta$ by linear regression, but instead of immediately setting $\hat{\theta} = \theta$, we should have $\hat{\theta}$ incrementally step towards $\theta$ while giving TEGD a chance to influence the change. It's a sort of tracking error driven lowpass filter on the linear regression. Lyapunov analysis shows how critical this feature is to closed-loop stability.

Inspired by this mechanism, we propose a new technique that provides all the same benefits of concurrent learning, but for adaptive controllers using observer-side parameter estimation. As usual, we will have our stochastic observer produce estimates $\hat{\underline{x}} = [\hat{x}\ \hat{\theta}]^T$. However, the controller will no longer use the observer's $\hat{\theta}$ directly. Instead, it will use a vector of "controller parameters" $\hat{\theta}_c$ which evolve as,

$$\dot{\hat{\theta}}_c := \Gamma\left(Y^T(x)e + \gamma(\hat{\theta} - \hat{\theta}_c)\right)$$

The right-hand term causes $\hat{\theta}_c$ to evolve towards the current best-fit parameters $\hat{\theta}$ that the observer is generating. Meanwhile, the left-hand term can tend the evolution towards minimizing error just like it does in concurrent learning. The overall adaptation rate is governed by $\Gamma > 0$ while $\gamma > 0$ lets us balance the two motives.

Fundamentally, this *is* concurrent learning. The only difference is that we are using a stochastic observer for data selection and parameter regression (and of course state estimation). This "stochastic concurrent learning" (SCL) technique gets the best of both observer-side parameter estimation and ordinary concurrent learning (OCL). The following list summarizes the advantages:

- Parameter estimation benefits from probabilistic state information.

- State estimation benefits from having an increasingly accurate process model.

- The whole algorithm is Markovian; no need to store or manage a history stack. (Technically, the stochastic observer *remembers everything* but weights those memories by their probabilistic likelihood of being erroneous and stores them in a single sufficient statistic).

- If the problem is linear-Gaussian, a simple Kalman filter can provide the maximum likelihood estimate at *all times*. (Other stochastic observers will heuristically approach this goal for the non-linear-Gaussian case).

- The more accurate, "pure" system identification $\hat{\theta}$ is still available to other programs, while the controller can independently use the more robust, "safe" system identification $\hat{\theta}_c$ that guarantees asymptotic stability of the tracking error.

- In addition to the usual belief distribution over the state-space, we will also have a belief distribution over $\theta$-space, which allows us to report error-bounds on our system identification.

- Intrinsically no need for state derivative measurements.

As the observer's $\underline{x}$ estimate converges (in the stochastic sense), we will have $\hat{\theta} \to \theta$, and SCL will behave exactly like OCL with a rich history stack. Notice that the

$$\sum_{i=1}^{N} Y_u^T(x_i, \dot{x}_i) Y_u(x_i, \dot{x}_i) > 0$$

condition of OCL is just a requirement that $\theta$ is (linearly) observable from the measurements we've obtained so far, so we actually haven't introduced any new conditions for SCL. For both SCL and OCL, all that's really happening is that $\Gamma Y^T(x) e$ is guaranteeing tracking error stability while we wait for linear regression to cause $\hat{\theta} \to \theta$. SCL just enables us to achieve $\hat{\theta} \to \theta$ more effectively.

**Marine ship demonstration:** We will now compare SCL to OCL by applying them to a realistic simulation of a marine ship. The simulation model (given in [31]) is an Euler-Lagrange dynamic with 6 states and 13 unknown inertial and drag parameters. It is highly nonlinear in the state, but linear in the parameters. This system was chosen because [7] uses OCL (specifically, integral concurrent learning) to tackle the same problem. The state and control input are,

$$x = \begin{bmatrix} \text{world\_x\_position} \\ \text{world\_y\_position} \\ \text{heading\_angle} \\ \text{surge\_velocity} \\ \text{sway\_velocity} \\ \text{yaw\_velocity} \end{bmatrix} \quad u = \begin{bmatrix} \text{surge\_force} \\ \text{sway\_force} \\ \text{yaw\_torque} \end{bmatrix}$$

and the dynamics are,

$$\dot{x} = \begin{bmatrix} Rv \\ M^{-1}\Big(u + (D - C)v\Big) \end{bmatrix}$$

where,
$$v = \begin{bmatrix} x_4 \\ x_5 \\ x_6 \end{bmatrix}$$

$$M = \begin{bmatrix} \theta_1 & 0 & 0 \\ 0 & \theta_2 & \theta_3 \\ 0 & \theta_3 & \theta_4 \end{bmatrix} > 0$$

$$C(x) = \begin{bmatrix} 0 & 0 & -(\theta_3 x_6 + \theta_2 x_5) \\ 0 & 0 & \theta_1 x_4 \\ \theta_3 x_6 + \theta_2 x_5 & -\theta_1 x_4 & 0 \end{bmatrix} = -C(x)^T$$

$$D(x) = \begin{bmatrix} \theta_5 |x_4| & 0 & 0 \\ 0 & \theta_6|x_5| + \theta_9|x_6| & \theta_{10}|x_5| + \theta_8|x_6| \\ 0 & \theta_{11}|x_5| + \theta_{12}|x_6| & \theta_{13}|x_5| + \theta_7|x_6| \end{bmatrix} < 0$$

$$R(x) = \begin{bmatrix} \cos(x_3) & -\sin(x_3) & 0 \\ \sin(x_3) & \cos(x_3) & 0 \\ 0 & 0 & 1 \end{bmatrix} = R^{-1^T}$$

We can also inject an external disturbance through $u$ by letting $u = u_{\text{control}} + u_{\text{disturb}}$.

The results in [7] assume exact state knowledge, so we will allow that in our first few simulations here too. Afterward we will examine simulations where only noisy measurements of the state are available (via some noisy "state sensor").

We will have the boat track a figure-8 pattern for most of the simulations. However, driving with heading tangent to the figure-8 (like one would expect a boat to do) does not excite the sway velocity state, so we will have the boat rotate about its center as it performs the figure-8.

Being as TEGD underlies any concurrent learning design, let's begin by looking at the performance of a pure TEGD adaptive controller. In Figure 7-1 (at the end of this section) we see that trajectory tracking is very good (although not perfect). In Figure 7-2, we see that the parameter estimates are not a valid system identification. This is expected for TEGD, which only moves the parameters to reduce instantaneous tracking error.

Next we will examine OCL on the same problem. We kept 100 data points in the history stack and used the minimum singular-value raising technique described in [32] for data selection. The integral-filter method of integral concurrent learning was used to avoid the need for $\dot{x}$ measurements (in comparison, SCL intrinsically never needs $\dot{x}$ measurements). In Figure 7-3 we see that tracking performance is basically perfect now, and in Figure 7-4 we see that the parameter estimates cleanly converged to their true values. This is unsurprising, since all the assumptions of OCL have been met exactly in this simulation.

Now let's take a look at SCL under the same conditions. I.e., our sensor model is exact state knowledge,

$$\underline{h}(\underline{x}, u, t, \nu) = x$$

leaving only the parameter part of $\underline{x}$ to be estimated. Since the system is linear in the parameters, all we have to do is guess a Gaussian prior on $\hat{\theta}$ for the Kalman filter to be the optimal observer. We will still employ a UKF because it is already coded up from the motor demonstration, but note that the UKF (and EKF) do exactly reduce to the basic Kalman filter in this setting. For our initial covariance, we set all the diagonals for $\hat{x}$ to $10^{-10}$ and all the diagonals for $\hat{\theta}$ to $9000$ to capture our initial uncertainty.

In Figure 7-5 we see the same perfect tracking we obtained with OCL, but in Figure 7-6 we see substantially faster parameter convergence (notice that this simulation was half as long). The finite history stack of OCL just can't compete with the all-remembering nature of a stochastic observer. Lastly, we see that the control parameters $\hat{\theta}_c$ follow the UKF parameters very closely after the first few seconds. In those first few seconds, $\hat{\theta}$ is far from $\theta$, so the slight initial tracking error causes $\hat{\theta}_c$ to deviate from $\hat{\theta}$ to maintain stability. I.e. we have all the same benefits of the OCL robustification term without any influence on our UKF's normal operation.

Figure 7-7 shows the parameter estimation error decay juxtaposed against the decay of the covariance diagonals. This is something we didn't have with OCL– a direct estimate of our uncertainty in the current parameters. The inertial and axis-aligned drag parameter estimates converged extremely quickly, which is reflected in their UKF-estimated variances rapidly dropping. The cross-flow drag parameters took a little longer (they are more difficult to excite) but their variances indicate that to us. The plot even shows us that around 7.5 seconds in, we transitioned through some states that made the remaining unknown parameters as observable as the inertial parameters were at the start. Being able to report a system identification covariance is very useful because in the real world we won't know exactly how wrong we are.

So what if we don't have exact state knowledge? Let's add Gaussian noise to our state sensor,

$$\underline{h}(\underline{x}, u, t, \nu) = x + \nu$$

$$\nu \sim \mathcal{N}(0, \text{ diag}(0.0025,\ 0.0025,\ 0.6,\ 0.0025,\ 0.0025,\ 4))$$

We placed most of the noise on the yaw velocity estimate, giving it a standard deviation of 2 degrees per second (imagine we couldn't afford a nicer gyroscope). We would expect, then, that parameter estimates closely associated with the yaw velocity state would be most affected. Also, note that none of the controller gains were changed between the noise-free simulations and the noisy simulations (for both OCL and SCL).

Figure 7-8 and 7-9 show the effect this has on OCL. The noise infects the history stack and has a detrimental effect on the estimate of $\theta_4$, our yaw moment of inertia. Almost all of our $\theta_4$ observability comes from the very start of the simulation when our yaw acceleration is largest. This information is now erroneous, but OCL holds onto it and begins pushing $\hat{\theta}_4$ towards a completely wrong value. The reason the yaw-related drag coefficients weren't as affected was because our constantly-rotating trajectory provided so much data for them that even an unweighted linear regression was able to filter out the noise. The other parameters are coupled to states with barely any noise, so their estimation behavior hasn't changed much.

Meanwhile, Figure 7-10 and 7-11 show that SCL handles the noise just fine. The reason is twofold:

1. Incorporating knowledge of the state sensor's covariance allows parameter estimation to be more cautious about erroneous data.
2. Increasingly accurate parameter estimates make our observer's predict more effective at filtering out sensor noise.

These phenomena are synergistic and greatly improve estimation (and consequently tracking). In Figure 7-10, notice how the noise in the yaw rate estimate is reduced with time– the observer's predict is getting better thanks to coupled parameter estimation. Meanwhile, the increase in state estimation accuracy ends up further helping parameter estimation. Figure 7-11 shows that, while not quite as fast as the no-noise case, the UKF still got us perfect parameter estimates.

In Figure 7-11, there is clearly a significant difference between $\hat{\theta}$ and $\hat{\theta}_c$ for the time before $\hat{\theta} = \theta$. The UKF, which only cares about system identification, sharply changes its estimates as it gets more information. Meanwhile, our controller uses a lowpass filtered version of the UKF estimates that gives TEGD a chance to increase robustness.

Finally, to really show off the power of SCL, we switch the desired trajectory to something more intense, add a surprise step disturbance in the middle of the run, and most importantly, add 3 more parameters to the UKF's process model so it can attempt to fit the disturbance itself. The results are shown in Figure 7-12 and 7-13. The UKF is able to excellently distinguish between what is a disturbance and what are the other modes of the model. The intense desired trajectory actually helped excite more modes more quickly and led to even faster convergence than we had with the figure-8.

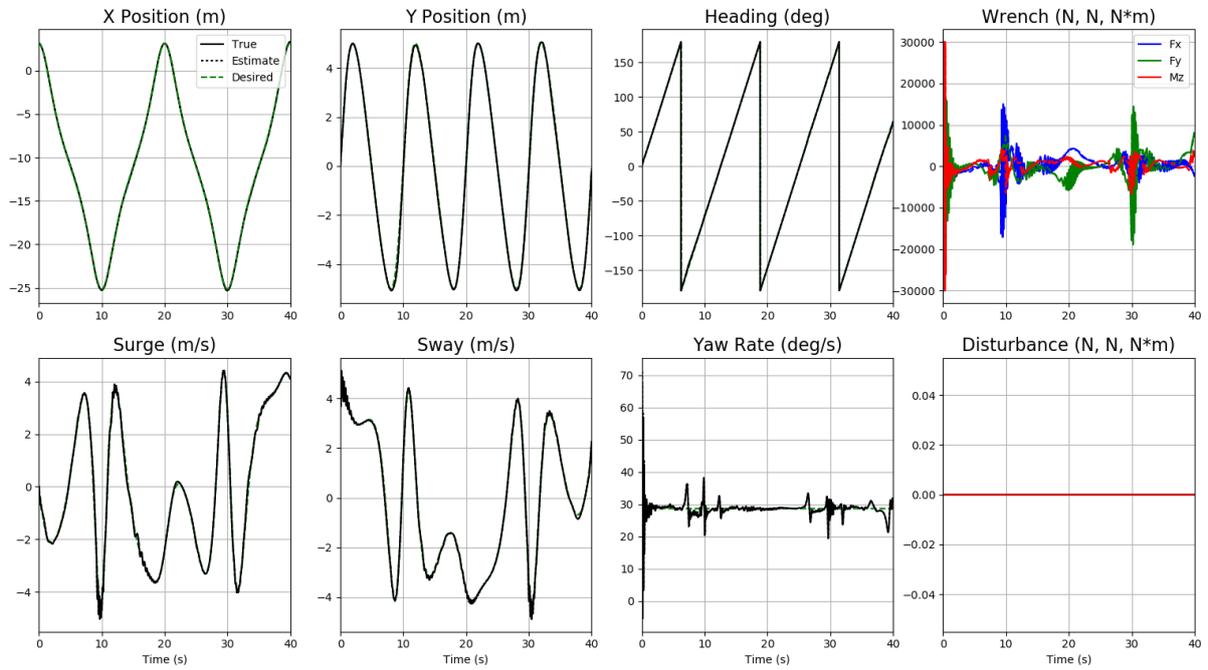

Figure 7-1. TEGD - state evolution.

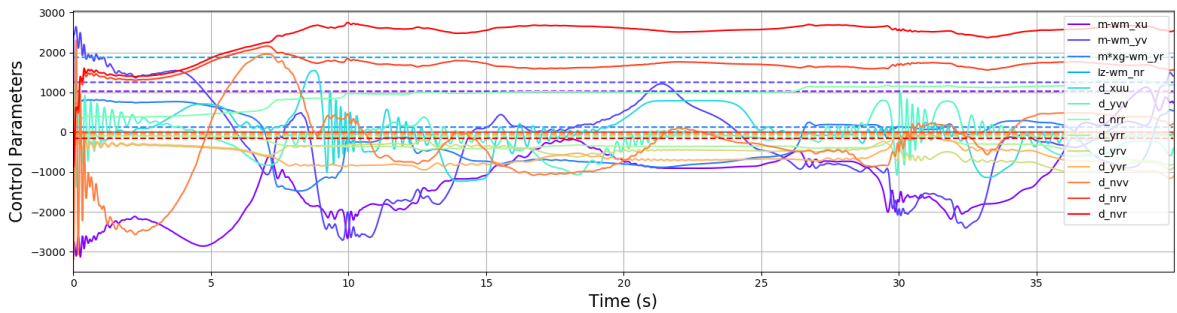

Figure 7-2. TEGD - parameter estimate evolution.

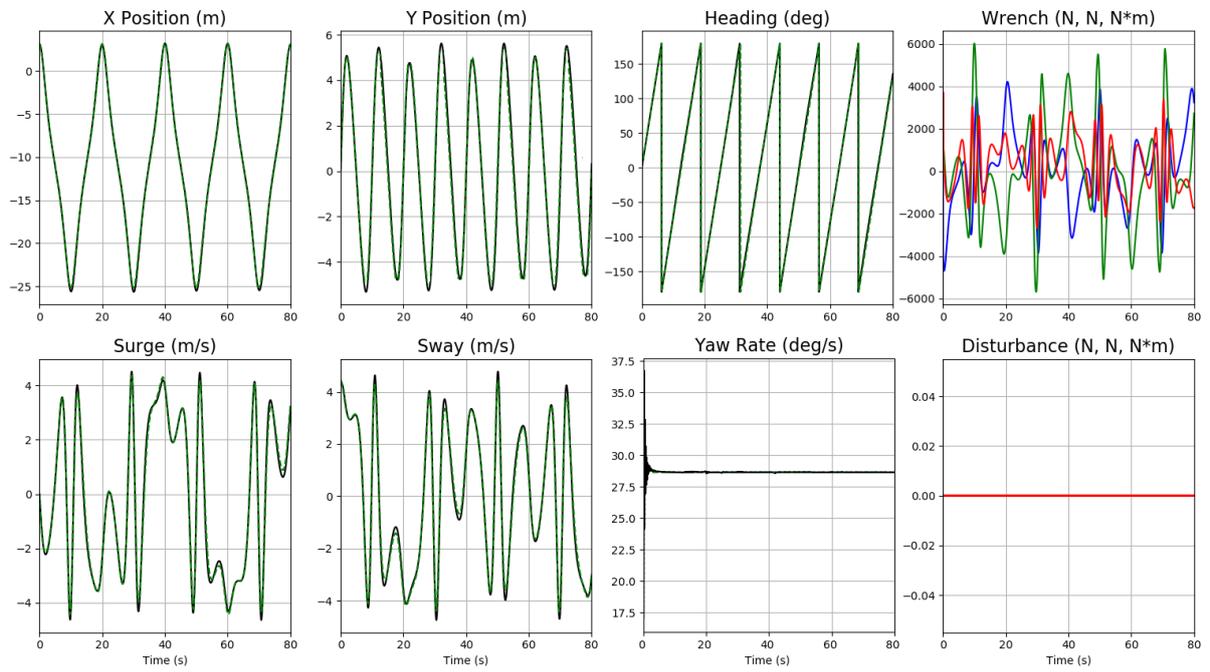

Figure 7-3. OCL - state evolution.

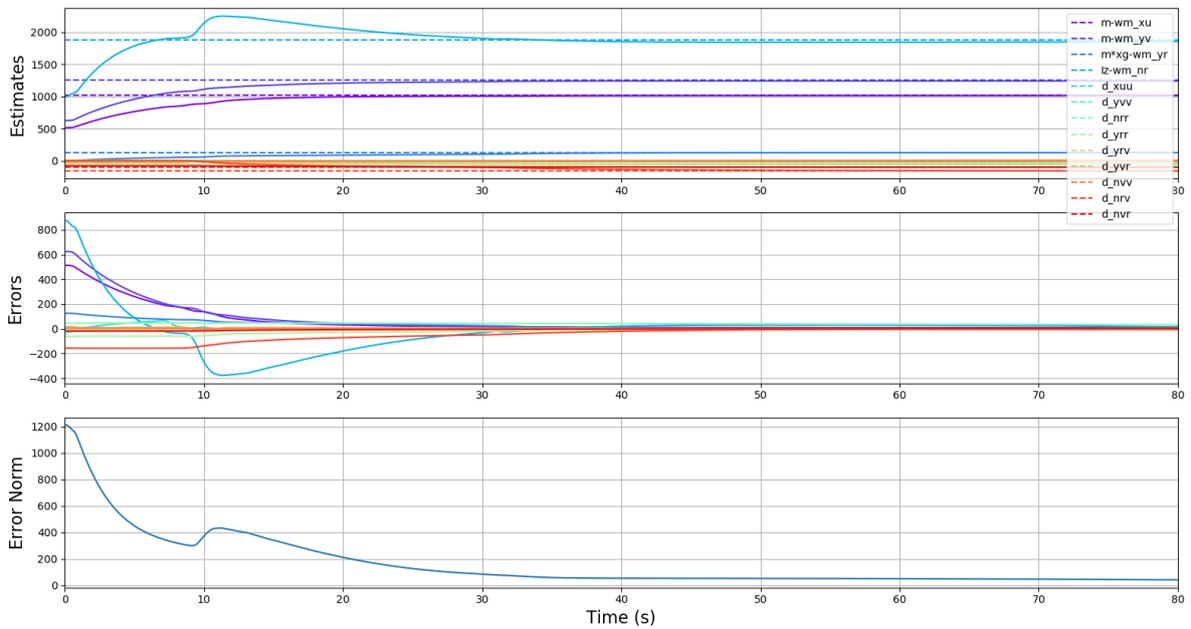

Figure 7-4. OCL - parameter estimate evolution.

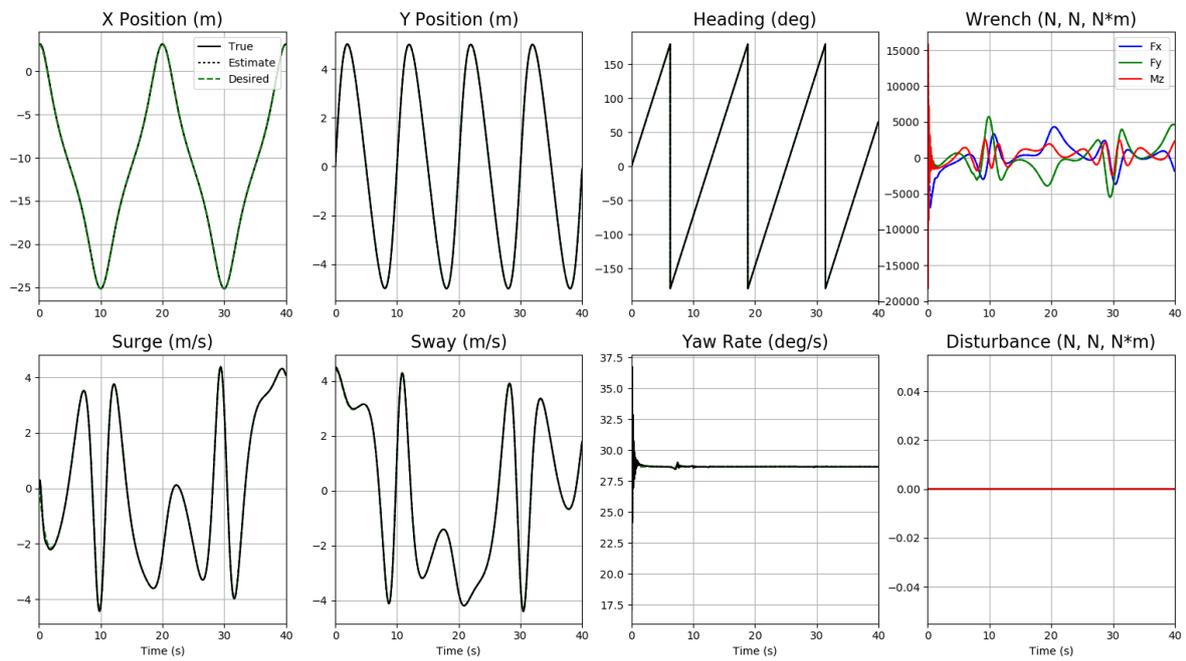

Figure 7-5. SCL - state evolution.

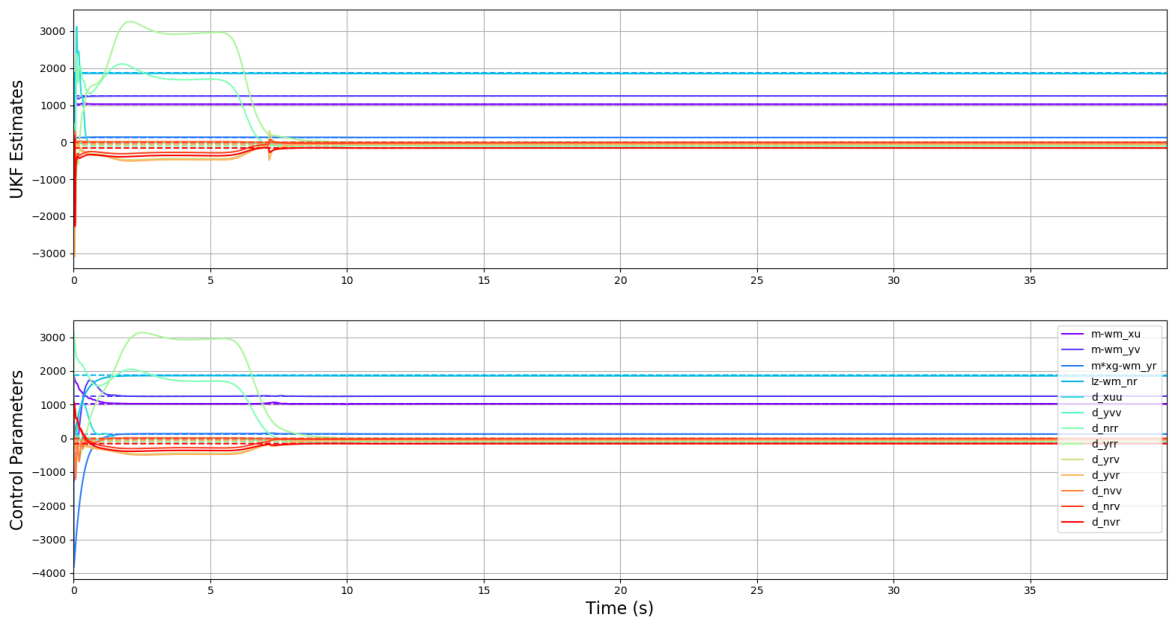

Figure 7-6. SCL - parameter estimate evolutions.

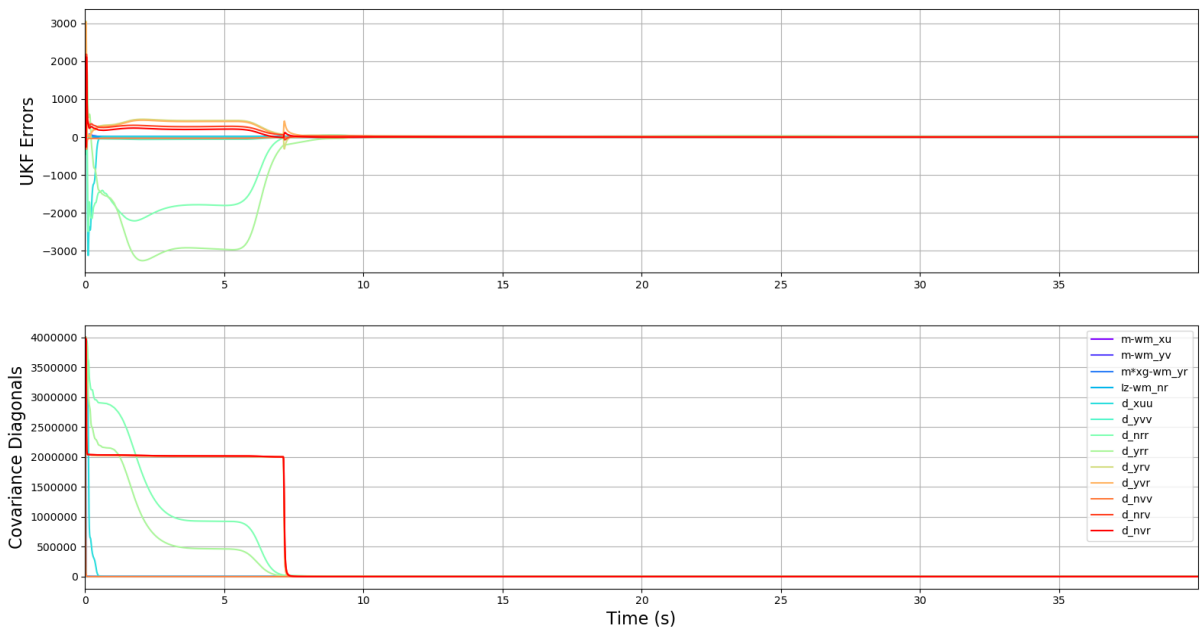

Figure 7-7. SCL - parameter estimate error and covariance evolutions.

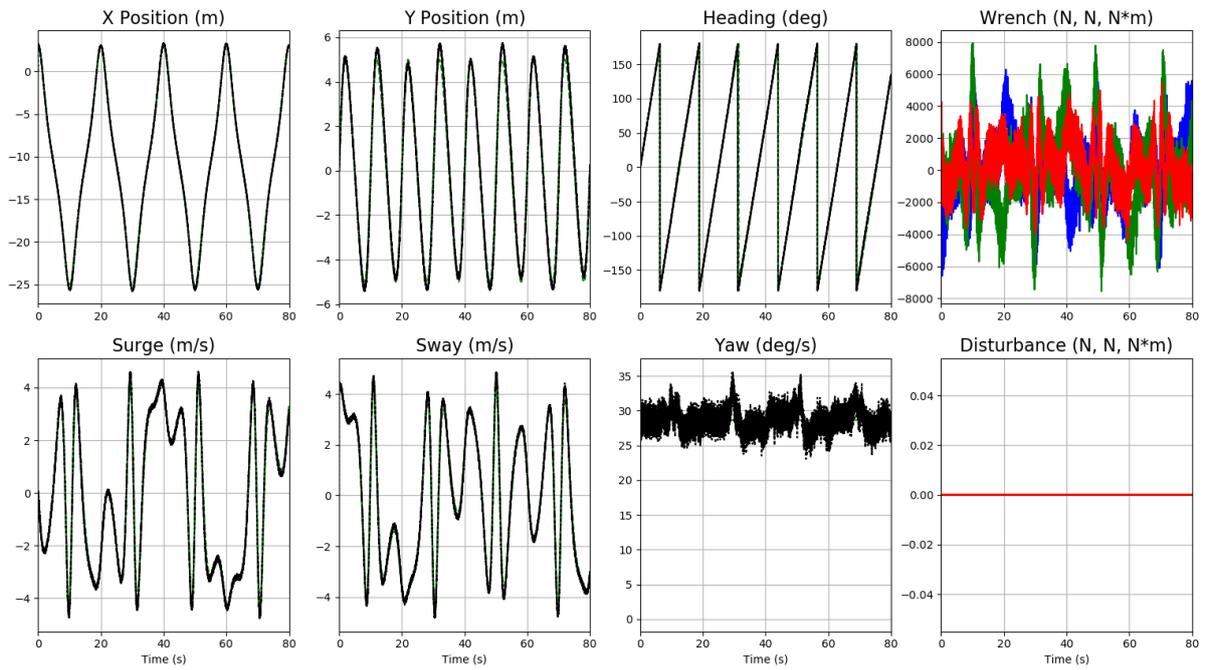

Figure 7-8. OCL against gyro noise - state evolution.

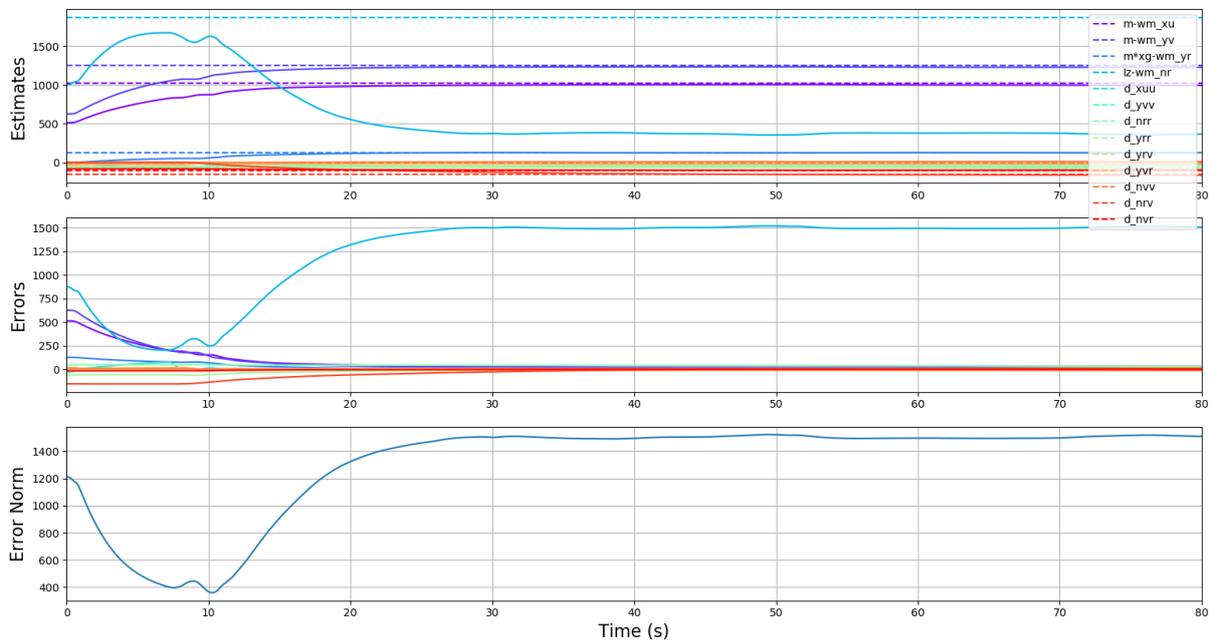

Figure 7-9. OCL against gyro noise - parameter estimate evolution.

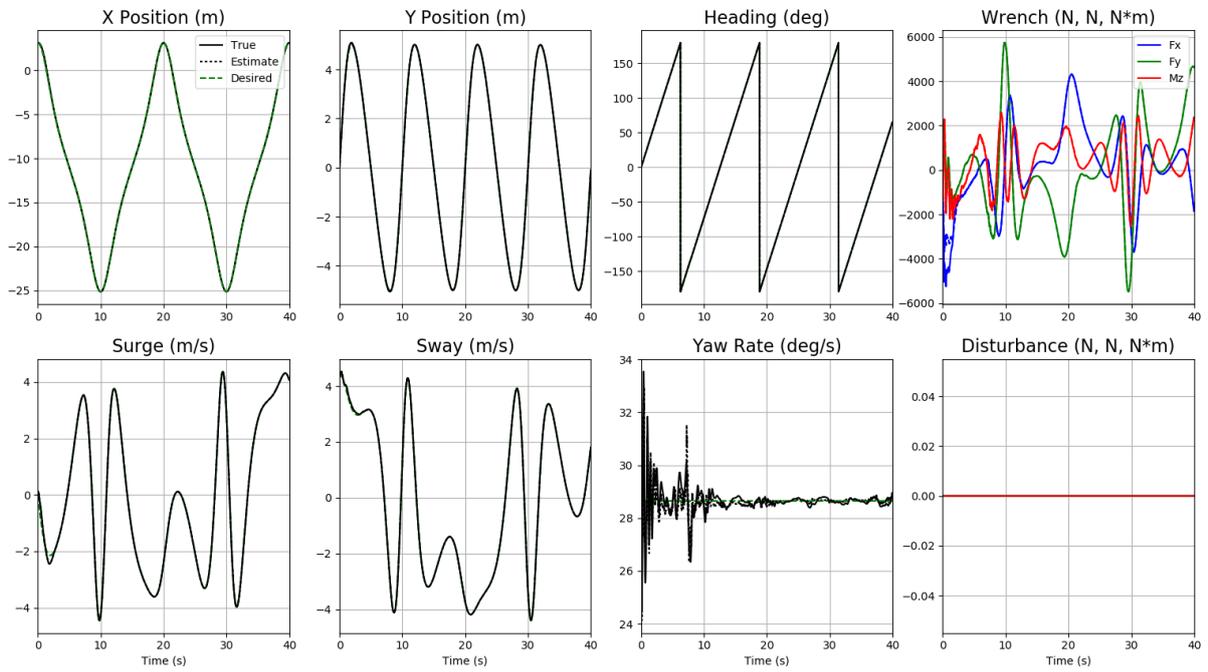

Figure 7-10. SCL against gyro noise - state evolution.

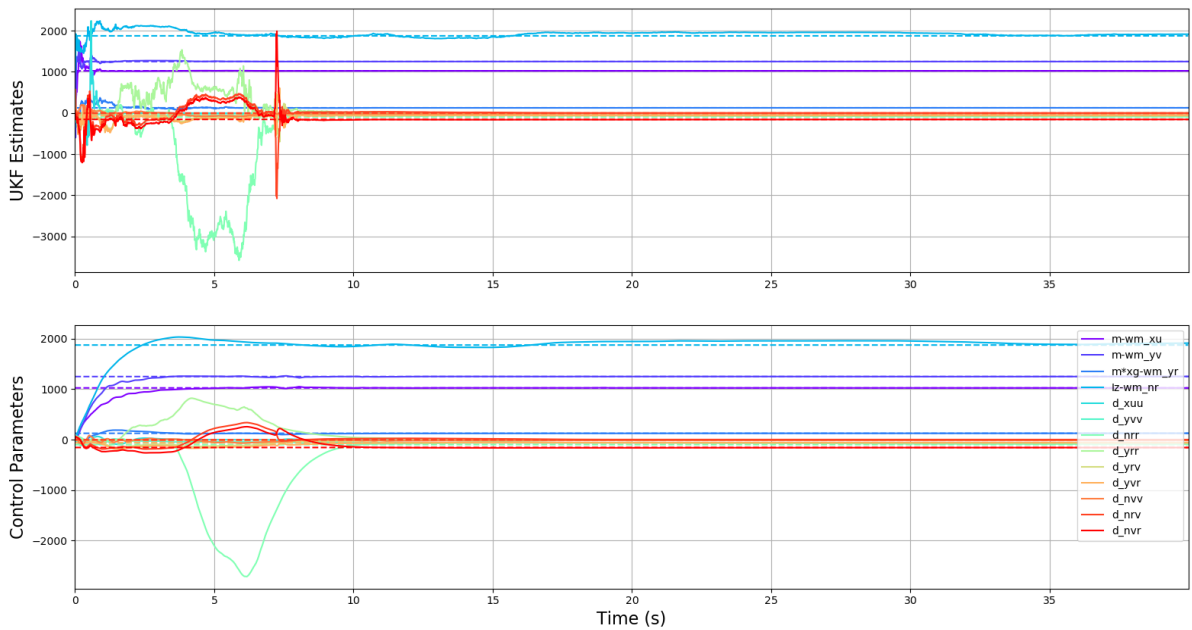

Figure 7-11. SCL against gyro noise - parameter estimate evolution.

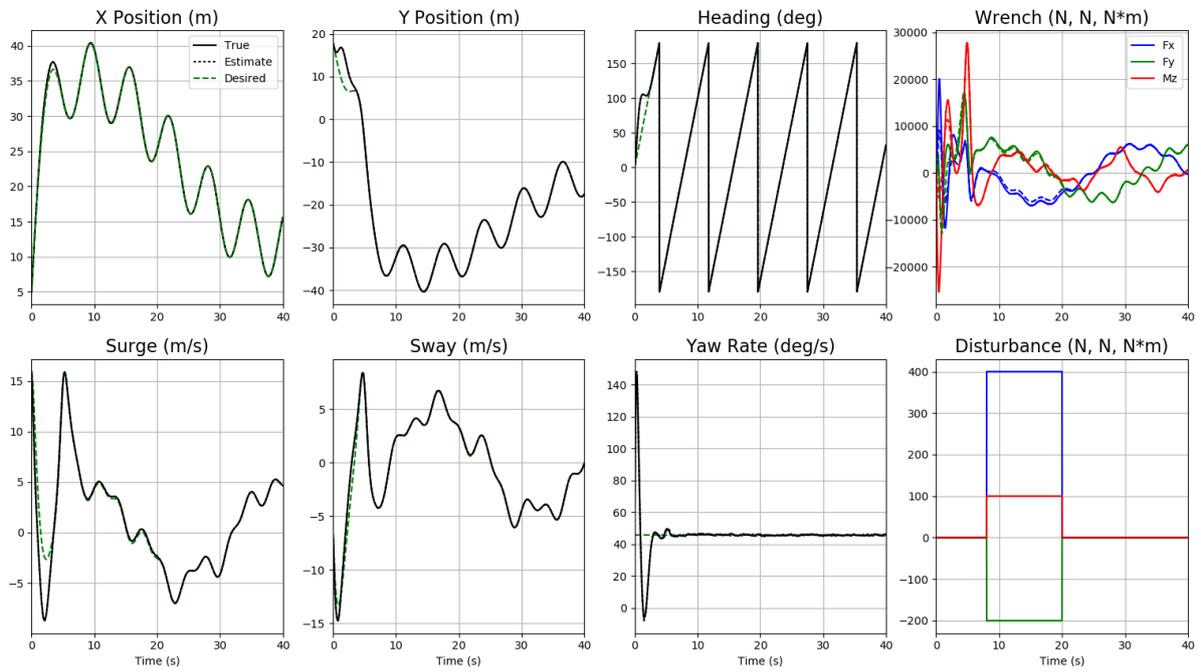

Figure 7-12. SCL against gyro noise and disturbance, with a more intense desired trajectory - state evolution.

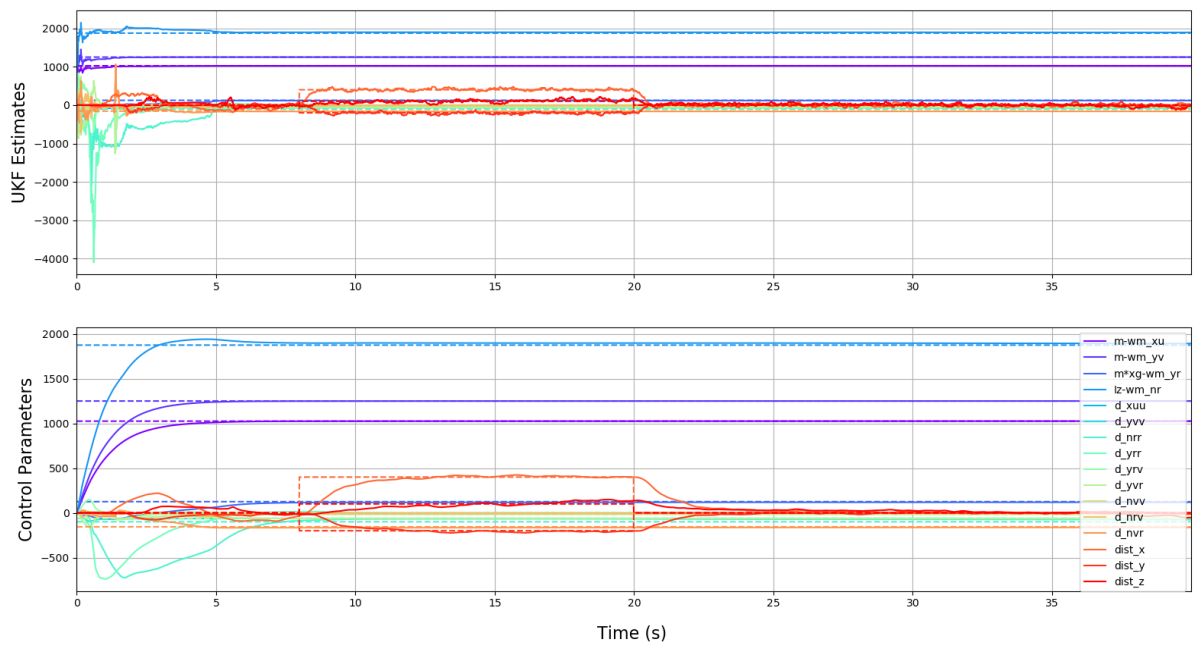

Figure 7-13. SCL against gyro noise and disturbance, with a more intense desired trajectory - parameter estimate evolution.

# CHAPTER 8
# CONCLUSION

It is not news that the powerful tools of stochastic estimation theory can be used for both state estimation *and* system identification. In this thesis, we additionally made clear just how beneficial connecting them can be. Therefore, it makes little sense to isolate state estimation and system identification from each other like many adaptive control designs do. Combining them within the framework of a stochastic observer is the obvious reconciliation. However, doing so can drown stability proofs in complexity.

Rather than fighting with this complexity, we proposed a new method that circumvents it. Instead of having the controller use the observer's potentially unsafe parameter estimates directly, we have it use its own "controller parameters" that only follow the observer's estimates in the absence of tracking error. If tracking error increases, a TEGD term will dominate the controller parameter evolution and tend it towards a stabilizing solution rather than the observer's solution. Thus, the controller is inherently safer during observer transients, while all the benefits of accurate observer-side parameter estimation are retained.

Viewing the observer as "probabilistic data selection and regression" reveals that this methodology is essentially just a stochastic extension to the logic of concurrent learning. We saw in our marine ship simulations that this "stochastic concurrent learning" outperforms ordinary history-stack-based concurrent learning in every way. We hope that this thesis acts as a seed for future research on our idea.

BIOGRAPHICAL SKETCH

Jason Nezvadovitz graduated summa cum laude from the University of Florida in 2017 with a Master of Science degree in dynamics and control theory, and a Bachelor of Science degree in mechanical engineering with minor in electrical engineering. As a student, he designed, manufactured, and programmed a wide variety of robotic systems including an autonomous submarine, pontoon boat, and mobile 4-DOF manipulator. These systems provided him with real-world platforms for his more theoretical research in the Nonlinear Controls and Robotics Lab at UF. His research surrounded adaptive control, estimation theory, and motion planning.